\documentclass[aip,apl,graphicx,%
amsmath,amssymb,%
reprint,%
]{revtex4-1}

\usepackage{graphicx}
\usepackage{dcolumn}
\usepackage{bm}
\usepackage[utf8]{inputenc}
\usepackage[T1]{fontenc}
\usepackage{hyperref}
\usepackage{amsmath}
\usepackage{upgreek}
\usepackage{hhline}
\usepackage{textcomp,gensymb}
\usepackage{blindtext}
\usepackage{color}
\usepackage{nicefrac}
\usepackage{bm}
\usepackage[]{changes}

\draft 

\begin{document}

\title{Characterization of plasmas driven by laser wavelengths in the $ 1.064 -  10.6\,\micron$ range as future extreme ultraviolet light sources} 

\newcommand{\ARCNL}{Advanced Research Center for Nanolithography,\\Science Park 106, 1098 XG Amsterdam, The Netherlands}
\newcommand{\VU}{Department of Physics and Astronomy, and LaserLaB, Vrije Universiteit Amsterdam,\\De Boelelaan 1081, 1081 HV Amsterdam, The Netherlands}

\author{D. J. Hemminga}
\affiliation{\ARCNL}
\affiliation{\VU}

\author{O. O. Versolato}

\affiliation{\ARCNL}
\affiliation{\VU}

\author{J. Sheil}

\affiliation{\ARCNL}
\affiliation{\VU}
\email{j.sheil@arcnl.nl}

\date{\today}

\newcommand{\Plas}{P_{\text{las}}}
\newcommand{\Pabs}{P_{\text{abs}}}
\newcommand{\Prfl}{P_{\text{ref}}}
\newcommand{\Pesc}{P_{\text{esc}}}
\newcommand{\Pref}{\Prfl}
\newcommand{\Poff}{\Pesc} 

\newcommand{\Pkin}{P_{\text{kin}}}
\newcommand{\Prad}{P_{\text{rad}}}
\newcommand{\Pinb}{P_{\text{inb}}}
\newcommand{\Pint}{P_{\text{int}}}

\newcommand{\abslasratio}{\Pabs/\Plas}
\newcommand{\rfllasratio}{\Prfl/\Plas}
\newcommand{\offlasratio}{\Poff/\Plas}

\newcommand{\kinabsratio}{\Pkin/\Pabs}
\newcommand{\radabsratio}{\Prad/\Pabs}
\newcommand{\inbabsratio}{\Pinb/\Pabs}
\newcommand{\intabsratio}{\Pint/\Pabs}

\newcommand{\CE}{\text{CE}_{\text{P}}}
\newcommand{\SP}{\text{SP}_{\text{P}}}
\newcommand{\SPratio}{\Pinb/\Prad}
\newcommand{\fracinbtwopi}{f_{\text{inb},2\pi}}
\newcommand{\CEratio}{\fracinbtwopi \Pinb/\Plas}

\newcommand{\micron}{\upmu\text{m}}
\newcommand{\llaser}{\lambda_{\text{laser}}}

\begin{abstract}
We characterize the properties of extreme ultraviolet (EUV) light source plasmas driven by laser wavelengths in the $\llaser = 1.064 - 10.6\,\micron$ range. Detailed numerical simulations of laser-irradiated spherical tin microdroplet targets reveal a strong laser-wavelength dependence on laser light absorptivity and the conversion efficiency of generating EUV radiation. 
Radiative losses are found to dominate the power balance for all laser wavelengths, and a clear shift from kinetic to in-band radiative losses with increasing laser wavelength is identified. 
We find that the existence of maximum conversion efficiency, near $ \llaser = 4\,\micron $, originates from the interplay between the optical depths of the laser light and the in-band EUV photons for this specific target geometry. 
\end{abstract}

\maketitle


 Extreme ultraviolet lithography (EUVL) is driving mass-production of today's most advanced integrated circuits (ICs) \cite{Moore2018,Bakshi2018}. 
 Crucial to the success of this technology has been the development of a sufficiently powerful, stable and ``clean'' source of EUV radiation \cite{Fomenkov2017} concentrated in a narrow $ 13.5 $\,nm $ \pm $ $ 1 \% $ region where molybdenum/silicon multilayer mirrors 
 exhibit high reflectance (the so-called ``in-band'' region) \cite{Bajt2002,Huang2017}. This radiation is most efficiently generated in a laser-produced plasma (LPP) formed when high-intensity CO$_{2}$ laser light (laser wavelength $ \llaser = 10.6\,\micron$) is focussed onto pre-shaped tin microdroplet targets \cite{Fujioka2008,Nishihara2008,Fujimoto2012,Kurilovich2016,Grigoryev2018,HernandezRueda2022,Versolato2019}. This plasma contains large populations of Sn$^{11+}$ $ - $ Sn$^{15+}$ ions which generate intense, narrowband EUV radiation through bound-bound atomic transitions\cite{Svendsen1994, Churilov2006b, Fujioka2005a, Sasaki2010, OSullivan2015, Scheers2020, Torretti2020}. 

Nowadays, industrial EUV sources generate a remarkable 250\,W of in-band EUV power\cite{Fomenkov2018}. Efforts to increase source powers beyond 600\,W are now underway to facilitate increased wafer throughput \cite{Brandt2022}. While CO$_{2}$ laser-driven plasmas are the backbone of current EUV sources for high-volume manufacturing \cite{Nowak2013,Fomenkov2017}, 
rapid developments \cite{Tamer2021} in solid-state lasers (which typically operate in the near- to mid-infrared wavelength range) 
make them a viable alternative \cite{Sizyuk2020,Schupp2021,Yuan2021} in the future due to their high efficiencies in converting electrical power to laser light and their potential for scaling to high average powers \cite{Siders2019, Tamer2021}. 
A crucial topic to be addressed in such studies is the impact of laser wavelength $\llaser$ on the radiative and kinetic properties of the plasma. The latter is especially important in the context of debris generation \cite{Fujioka2005,Hemminga2021} which may limit the lifetime of optical components. 
The laser wavelength sets the critical electron density (the density at which the plasma becomes opaque to incident laser radiation) according to $ n_{\mathrm{e,cr}} \propto \llaser^{-2}$. This determines the densities for which EUV radiation is generated and must propagate through. High plasma densities or long path lengths generate large optical depths which redistribute in-band energy into other channels, e.g.,  via spectral broadening \cite{Schupp2019a}. 
While this effect can be negated by moving to long laser wavelengths \cite{Torretti2019}, the absorption path length of laser photons decreases with increasing $ \llaser$. In the presence of steep electron density gradients, this can lead to significant laser reflection from the critical surface and a loss of input laser energy \cite{Basko2015,Basko2016}. Efficiency at different laser wavelengths is therefore determined by a trade-off in numerous underlying physical processes. The goal of the present work is to quantify this trade-off for laser wavelengths lying in the largely unexplored region between $ \llaser = 1.064\,\micron $ (Nd:YAG laser) and $ 10.6\,\micron$ although the existence of an optimum laser wavelength has been suggested previously in the EUV Source community \cite{Langer2019,Fomenkov2019}.

In this letter, we present a comprehensive characterization of the properties of EUV source plasmas driven by laser wavelengths in the $ 1.064 \leq \llaser \leq 10.6\,\micron$ range. 
We identify a strong laser-wavelength dependence on laser absorptivity and EUV generation efficiency, and, moreover, report on the early-time establishment of steady-state plasma flows. The partitioning of laser power into plasma components (kinetics, internal and radiated power) is quantified and compared with the predictions of analytical theory \cite{Nishihara2006,Murakami2006}. Finally, we delve into the factors underlying a maximum in the conversion efficiency (ratio of in-band power radiated into the laser-illuminated half-sphere to laser power) for $ \llaser \approx 4\,\micron $ laser irradiation. 

\begin{figure*}
\includegraphics[width=\textwidth]{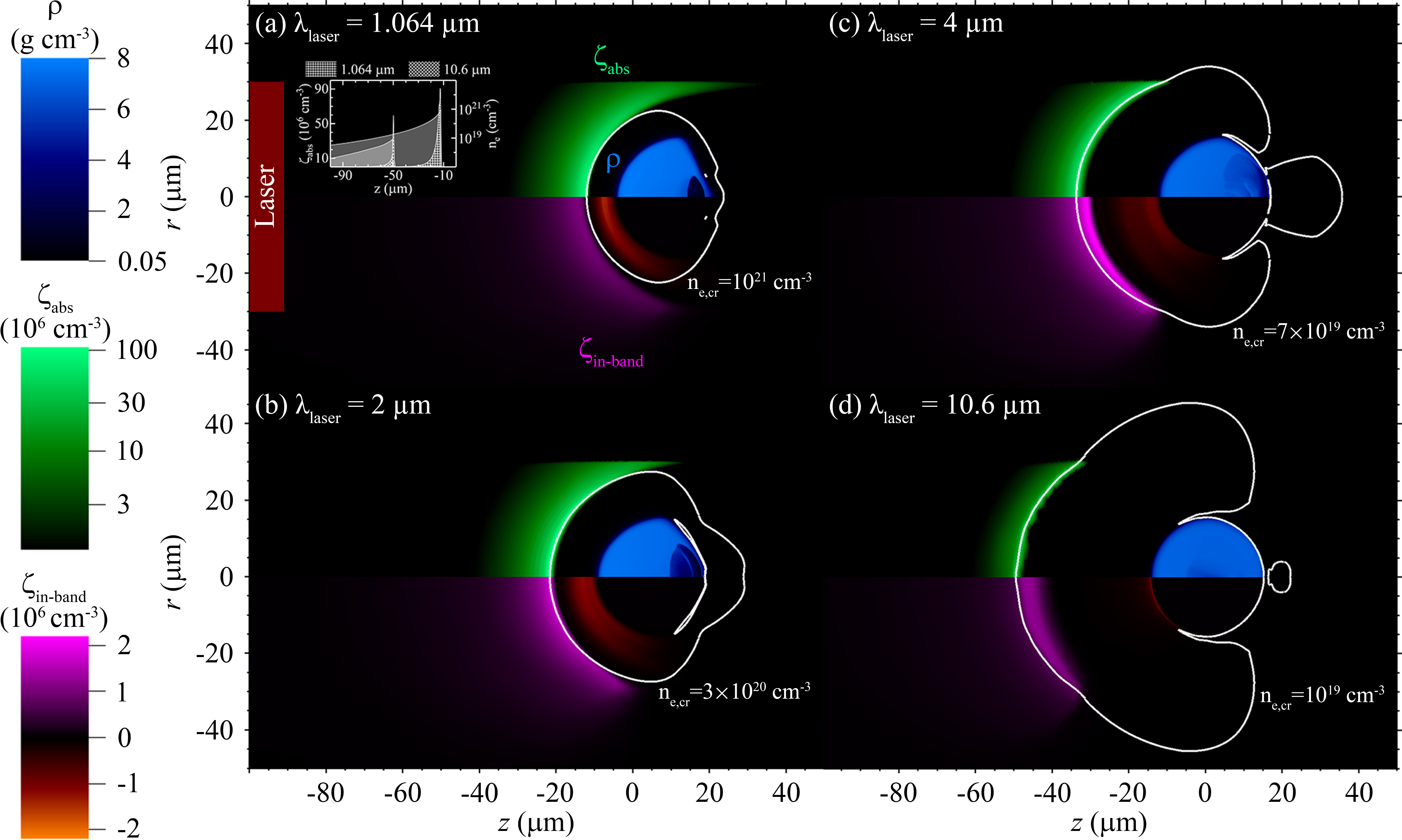}%
\caption{\label{fig:pseudocolor}%
The instantaneous (at $ t = 18 $ ns) volume-specific laser deposition rate \cite{Basko2017a} $\zeta_{\text{abs}}$ (green) and in-band radiative power $\zeta_{\text{in-band}}$ (purple/orange) normalized by the input laser power for $\llaser=$ (a) 1.064, (b) 2, (c) 4 and (d) 10.6\,$\micron$. The critical electron density (white contour) and fluid density $ \rho $ (blue) are indicated. The inset in (a) depicts the electron density and $\zeta_{\text{abs}}$ along the laser axis for $\llaser=$ 1.064 (square meshing) and 10.6\,$\micron$ (diamond meshing).
}
\end{figure*}

We have performed numerical simulations of laser-produced tin plasmas using the two-dimensional radiation-hydrodynamic code RALEF-2D \cite{Basko2010,Basko2012,Tauschwitz2013} tailored for EUV source applications \cite{Basko2016,Basko2017b,Kurilovich2018,Hemminga2021,HernandezRueda2022}. In brief, the code solves the single-fluid, single-temperature hydrodynamic equations incorporating radiation transfer and thermal conduction processes \cite{Basko2009}. Spectral absorption coefficients and equation-of-state parameters are derived from the THERMOS code \cite{Nikiforov2005, Vichev2019} and the Frankfurt equation-of-state (FEOS) model\cite{Faik2018}, respectively. 
Laser light absorption and reflection is treated using a hybrid model combining a geometrical-optics ray-tracing approach in low-density plasma regions and a wave-optics approach in regions near and beyond the critical electron density\cite{Basko2017a}. 
Laser absorption coefficients are derived from the complex dielectric permittivity of the plasma as per the Drude model \cite{Elezier2002}. 
The simulated cases 
consider laser irradiation of 30-\,$\micron$-diameter liquid tin droplets, close to the industry standard, with spatially constant laser fluences of 60\,$\micron$ in width. The laser pulses are temporally trapezoidal-shaped with pulse lengths of 20\,ns (rise and fall times of $0.2$\,ns).
These experimental parameters are prototypical for recent simulation and experimental works alike (see, e.g., Refs.\,\onlinecite{Basko2015,Basko2016,Kurilovich2016,Schupp2019,Schupp2019a,Schupp2021,Schupp2021a,Shimada2005,Giovannini2013,Giovannini2014,Giovannini2015,Sizyuk2006}).
The laser wavelengths considered in this work are $\llaser = 1.064, 2, 3, 4, 5, 7, $ and $ 10.6\,\micron$. This encompasses two distinct regimes of laser absorption, where absorption occurs primarily in the (i) underdense corona (for small $ \llaser $) or (ii) a narrow region near the critical surface (for long $ \llaser $) \cite{Basko2015}.
The laser intensity is scaled according to $ I_{\mathrm{laser}} = (1.4\times 10^{11})/\llaser $~W/cm$^{2}$, 
an experimentally-motivated scaling which yields high conversion efficiencies for the laser wavelengths considered in this study \cite{Schupp2019,Behnke2021}. We note the close similarity between this scaling and the optimum laser intensity 
$ I_{\mathrm{laser}} \propto \llaser^{-1.2} $ proposed by Nishihara \textit{et al.} \cite{Nishihara2006}.  

In Fig.\,\ref{fig:pseudocolor}, we provide a still of the plasma formation induced by the four laser wavelengths $\llaser = 1.064, 2, 4, 10.6\,\micron$ at the time $ t = 18 $ ns after the laser pulse is switched on. 
The absorbed laser power per unit volume normalized by the input laser power, denoted $\zeta_{\text{abs}}$, is shown in the upper halves of the panels. We find that with increasing $ \llaser $, the laser absorption zone moves further away from the droplet (blue region) due to the inverse square dependence of $ n_{\mathrm{e,cr}} $ (white contours) on $ \llaser $ and the electron density gradient associated with the near-spherical flow \cite{Basko2016} $ n_{e} \propto r^{-2}$. Moreover, the spatial extent over which the laser light is absorbed reduces with increasing $ \llaser $, a direct result of the aforementioned transitioning between the two distinct regimes of laser absorption. This is exemplified in the inset of panel (a), which shows the electron density and $\zeta_{\text{abs}}$ along the laser axis for $\llaser=$ 1.064 (square meshing) and 10.6\,$\micron$ (diamond meshing). The steep electron density gradient precludes efficient absorption of long-wavelength laser light. In panel (d), we see that CO$_{2}$ laser absorption is restricted to a narrow region in front of the critical surface. Adopting Kramers cross section for inverse bremsstrahlung and assuming a (i) constant-temperature laser absorption zone and (ii) $ n_{e} \propto r ^{-2} $ profile, the optical depth of laser light $ \tau_{\mathrm{laser}}$ can be written \cite{Basko2015}

\begin{equation}
    \tau_{\mathrm{laser}} \approx 4 \times 10^{-2}  \frac{Z\ln(\Lambda)}{T^{3/2}\llaser}\left(\frac{R_{mc}}{a^{3}_{uv}\llaser}\right)
\end{equation}

\noindent where $ Z $, $ \ln(\Lambda) $, $ T $, $ R_{mc} $ and $ a_{uv} $ are the charge state, Coulomb logarithm, temperature (in hundreds of eV), radius of the critical surface and a dimensionless parameter determining the position of laser absorption. Inferring appropriate values from our simulations, we estimate $ \tau_{\mathrm{1.064 \micron}} \sim 10 $ and $ \tau_{\mathrm{10.6 \micron}} \sim 1 $ as orders of magnitude for the present illumination geometry.

In the bottom halves of Fig.\,\ref{fig:pseudocolor}, we show the net in-band radiated power per unit volume normalized by the input laser power, denoted $\zeta_{\text{in-band}}$. This provides a local measure of the efficiency of converting laser light to in-band radiation. Of the four cases shown, the 4$\,\micron$-driven plasma exhibits the highest $\zeta_{\text{in-band}}$ due to an optimum combination of intermediate laser absorptivity and intermediate optical depth of EUV photons (see Fig.\,\ref{fig:pseudocolor} and the discussion below). We see that with increasing $ \llaser $, the region of net in-band emission (purple regions) moves from regions with $ n_{\mathrm{e}} < n_{\mathrm{e,cr}} $ to regions with $ n_{\mathrm{e}} > n_{\mathrm{e,cr}} $. 
Furthermore, the regions of net absorption of in-band radiation (orange regions) are located close to the droplet (regions with high densities and low temperatures), which will improve material ablation from the droplet surface \cite{Nishihara1982,Nozomi2015}. 


\begin{figure}
\includegraphics{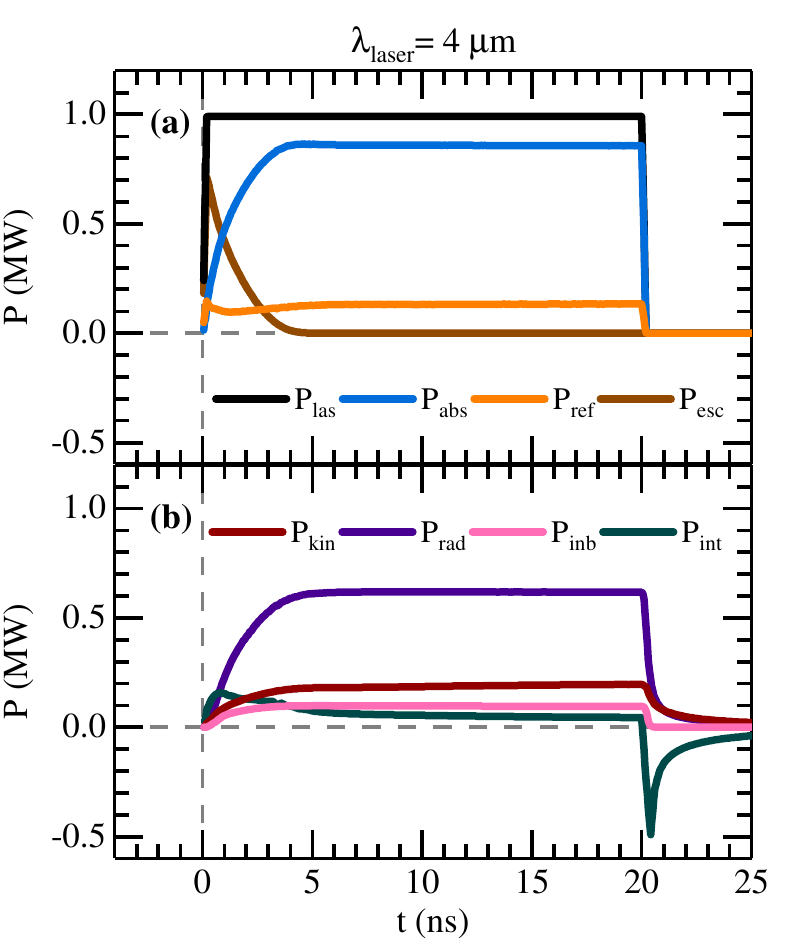}%
\caption{\label{fig:power-in-time} %
Time-dependent partitioning of laser power into (a) laser- and (b) plasma-based components for $ \llaser = 4\,\micron$ irradiation of a 30-\,$\micron$-diameter tin droplet.
}%
\end{figure}

\begin{figure}[ht!]
\includegraphics{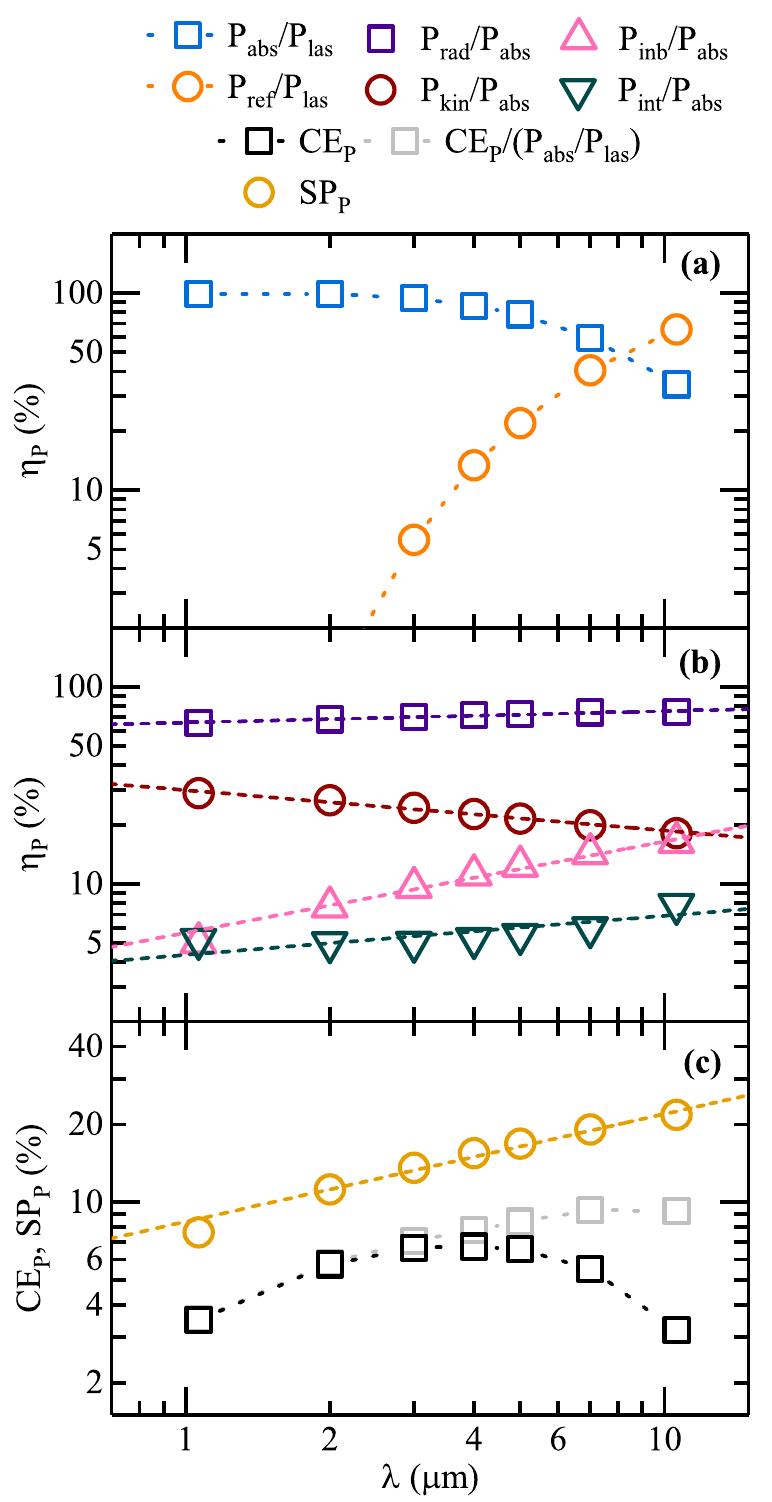}%
\caption{\label{fig:power-ratio} %
The instantaneous (at $ t = 18 $ ns) partitioning of (a) laser-based components (normalized by $ \Plas $) (b) plasma-based components (normalized by $ \Pabs $) and (c) spectral purity $ \mathrm{SP_{P}} $ (yellow circles), conversion efficiency $ \mathrm{CE_{P}} $ (black squares) and $ \mathrm{CE_{P}}/(\Pabs/\Plas) $ (gray squares)  as a function of laser wavelength. The dashed curves represent power-law fits to the data.
}%
\end{figure}

The instantaneous partitioning of laser power during $ \llaser = 4\,\micron $ illumination is shown in Fig.\,\ref{fig:power-in-time}. In Fig.\,\ref{fig:power-in-time} (a), we show the input $ \Plas $ (black), absorbed $ \Pabs $ (blue), reflected $ \Pref $ (orange) and ``escaped'' $ \Pesc $ (brown) laser power components. The escaped component represents laser radiation that initially misses the target, a quantity which decreases rapidly as the plasma expands and starts absorbing incident laser radiation. After 5 ns, a ``steady-state'' plasma flow regime is established whereafter $ \Pabs $ and $ \Pref $ attain near-constant values. This behaviour is evident in the plasma-based components $ \Pint $ (internal power $ - $ derived from the specific Helmholtz free energy\cite{Faik2014}, green), $ \Pkin $ (kinetic power, red), $ \Prad $ (total radiated power, purple) and $ \Pinb $ (in-band power, pink) shown in Fig.\,\ref{fig:power-in-time} (b). It is well-known that plasmas containing high-$Z$ ions exhibit large radiative losses \cite{Putterich2008}, and our simulations indicate that approximately 70\% of the absorbed laser power is channeled into radiation. Moreover, we find that nearly $ 16\% $ of this radiation is concentrated in the in-band region, a surprisingly large fraction given the narrowness (0.27\,nm) of this wavelength region. 
Finally, we quantify power partitioning as a function of laser wavelength. This enables a comprehensive characterization of EUV plasma source conditions, where high laser absorptivities coupled with large in-band radiative losses and minimal kinetic losses are most desired. 
In Fig.\,\ref{fig:power-ratio}(a), we present the ratios $ \Pabs/\Plas $ (blue) and $ \Pref/\Plas $ (orange) as a function of laser wavelength at steady-state conditions. The dashed curves represent power-law fits to the data.
We note the significant increase in laser reflectivity comparing $ \llaser = 1.064 $ and 10.6\,$\micron$ cases, which is due to the significant reduction of $ \tau_{\mathrm{laser}} $ with increasing $ \llaser $ specific to the current target geometry.


The internal, radiated and kinetic components exhibit their own unique dependencies on laser wavelength. In Fig.\,\ref{fig:power-ratio} (b), we show the ratios $ \Prad/\Pabs $ (purple squares), $ \Pkin/\Pabs $ (red circles), $ \Pinb/\Pabs $ (pink triangles), and $ \Pint/\Pabs $ (green inverted triangles). The origin of these power laws is not exactly known, and they most likely originate from a complex interplay of radiation-transport, laser absorption and plasma expansion effects. The 1D planar isothermal expansion model of Murakami \textit{et al.} \cite{Murakami2006}, for example, predicts that $ \Pkin/\Prad \propto \llaser^{-2}T^{-5/2} $, a much stronger dependence on $ \llaser $ than the $ \llaser^{-0.3} $ power law found in the present simulations. In addition to 2D expansion effects, this discrepancy may be attributed to the assumption in Murakami \textit{et al.} \cite{Murakami2006} that laser absorption occurs entirely at the critical surface, which is not true for short-wavelength lasers \cite{Basko2015,Schupp2021}.
With increasing $ \llaser $ (and therefore decreasing plasma density), the optical depth of EUV photons reduces from $ \tau_{\mathrm{EUV}} \approx 6 $ (Nd:YAG-driven plasma) to $ \tau_{\mathrm{EUV}} \approx 0.5 $ (CO$_{2}$-driven plasma) \cite{Schupp2019a}. This limits the degree of spectral broadening and redistribution of in-band energy into other channels, which explains the observed increase of $ \inbabsratio $ with increasing $ \llaser $ and the behaviour of the spectral purity $\SP = \Pinb/\Prad$ (defined in the full $4\pi$) presented in Fig.\,\ref{fig:power-ratio}(c). 
As the relative fraction of radiative losses increases with increasing $ \llaser $, the balance dictates a corresponding decrease of kinetic losses. 

The efficiency of producing in-band EUV radiation as a function of laser wavelength is shown in Fig.\,\ref{fig:power-ratio}(c). 
The conversion efficiency $\CE$ (black squares) exhibits a concave dependence on $ \llaser $ with a maximum at $ \llaser = 4\,\micron$. This maximum arises from the rather unique combination of laser absorptivity and $ \tau_{\mathrm{EUV}} $ values. 
In essence, the plasma conditions are in a ``sweet spot'' intermediate to the extreme cases of high laser absorptivity/low spectral efficiency ($ \llaser = 1.064 \,\micron $) and low absorptivity/high spectral efficiency ($ \llaser = 10.6 \,\micron $). This explains the simulation results of Langer \textit{et al.}, who identified an optimum for $ \llaser = 4.5$\,$\micron$ irradiation of a one-dimensional tin vapour target\cite{Langer2019}. 
It is worthwhile noting that the maximum is located on a rather flat part of the curve between 3 and 5\,$\micron$, and that the $\CE$ increase from 1.064\,$\micron$ to 2\,$\micron$ is rather substantial, in line with experimental observations \cite{Behnke2021}. 

The strong dependence of conversion efficiency on laser absorptivity for $ \llaser > 4 \,\micron $ substantiates the opportunity to improve $\CE$ for long laser wavelengths. In Fig.\,\ref{fig:power-ratio}(c), we plot the quantity $ \CE/(\abslasratio) $ (grey squares), which represents the conversion efficiency if the absorption fraction would be unity for all laser wavelengths. This quantity increases monotonically with increasing $ \llaser $, plateauing between $ \llaser = 7 $ and $ 10.6 \,\micron$. This behaviour can be understood qualitatively from the power balance model of Murakami \textit{et al.}\cite{Murakami2006}, where, denoting $ \eta_{\mathrm{C}} \equiv \CE/(\abslasratio) $, the model predicts that

\begin{equation}
    \eta_{\mathrm{C}} \approx 2g_{\mathrm{2D}}\left(1+\frac{5\Pkin}{2\Prad}\right)^{-1}\frac{S_{\mathrm{EUV}}}{\sigma T^{4}}
\end{equation}
\noindent where $ g_{\mathrm{2D}} $ is a factor accounting for 2D effects, $ S_{\mathrm{EUV}} $ is the in-band emissivity and $ \sigma $ is the Stefan-Boltzmann constant. Kinetic losses decrease with increasing laser wavelength. In fact, as $ \llaser \rightarrow \infty $, $ \Pkin/\Prad \propto \llaser^{-2}T^{-5/2}  \rightarrow 0 $ and $ \eta_{\mathrm{C}} $ approaches the constant value $ 2g_{\mathrm{2D}}S_{\mathrm{EUV}}/\sigma T^{4} $, behaviour which is consistent with the plateau observed in Fig.\,\ref{fig:power-ratio}(c). In order to increase laser absorptivity for long $ \llaser $, one could pre-irradiate the target to convert it into a rarefied, spatially extended medium. This would decrease the plasma density gradient and subsequently increase the laser optical depth and, thus, its absorption in the plasma. Such target pre-shaping has been successfully applied in industrial applications, enabling high conversion efficiencies from CO$_{2}$ laser-irradiated tin targets\cite{Fomenkov2017}. That said, target shaping remains unexplored in the intermediate wavelength region considered in this work, and this may lead to substantial increases in $ \CE $. 



In summary, we have investigated the power partitioning in a laser-produced tin plasma for laser wavelengths in the $ 1.064 \leq \llaser \leq 10.6 \,\micron $ range. 
We have identified a strong laser-wavelength dependence of laser absorptivity and the location of EUV generation. With increasing laser wavelength, the power balance monotonically shifts from kinetic losses to in-band radiative losses. The decrease in laser absorption for long laser wavelengths, combined with an concurrent decrease in EUV optical depth, yields a non-monotonic behaviour of the conversion efficiency, leading to an optimum at $ \llaser = 4\,\micron$. EUV sources based on long laser wavelengths would benefit from additional target preparation to ensure a higher absorption fraction.

\begin{acknowledgments}
We would like to thank Wim van der Zande for useful discussions. This project has received funding from European Research Council (ERC) Starting Grant number 802648. This work has been carried out at the Advanced Research Center for Nanolithography (ARCNL), a public-private partnership of the University of Amsterdam (UvA), the Vrije Universiteit Amsterdam (VU), the University of Groningen (RUG - associate partner), NWO and the semiconductor equipment manufacturer ASML. 
This work made use of the Dutch national e-infrastructure with the support of the SURF Cooperative using grant no. EINF-1043 and EINF-2947.
\end{acknowledgments}

\section*{References}
\bibliographystyle{apsrev4-1}
\bibliography{DH_APL2022}

\begin{thebibliography}{60}%
\makeatletter
\providecommand \@ifxundefined [1]{%
 \@ifx{#1\undefined}
}%
\providecommand \@ifnum [1]{%
 \ifnum #1\expandafter \@firstoftwo
 \else \expandafter \@secondoftwo
 \fi
}%
\providecommand \@ifx [1]{%
 \ifx #1\expandafter \@firstoftwo
 \else \expandafter \@secondoftwo
 \fi
}%
\providecommand \natexlab [1]{#1}%
\providecommand \enquote  [1]{``#1''}%
\providecommand \bibnamefont  [1]{#1}%
\providecommand \bibfnamefont [1]{#1}%
\providecommand \citenamefont [1]{#1}%
\providecommand \href@noop [0]{\@secondoftwo}%
\providecommand \href [0]{\begingroup \@sanitize@url \@href}%
\providecommand \@href[1]{\@@startlink{#1}\@@href}%
\providecommand \@@href[1]{\endgroup#1\@@endlink}%
\providecommand \@sanitize@url [0]{\catcode `\\12\catcode `\$12\catcode
  `\&12\catcode `\#12\catcode `\^12\catcode `\_12\catcode `\%12\relax}%
\providecommand \@@startlink[1]{}%
\providecommand \@@endlink[0]{}%
\providecommand \url  [0]{\begingroup\@sanitize@url \@url }%
\providecommand \@url [1]{\endgroup\@href {#1}{\urlprefix }}%
\providecommand \urlprefix  [0]{URL }%
\providecommand \Eprint [0]{\href }%
\providecommand \doibase [0]{http://dx.doi.org/}%
\providecommand \selectlanguage [0]{\@gobble}%
\providecommand \bibinfo  [0]{\@secondoftwo}%
\providecommand \bibfield  [0]{\@secondoftwo}%
\providecommand \translation [1]{[#1]}%
\providecommand \BibitemOpen [0]{}%
\providecommand \bibitemStop [0]{}%
\providecommand \bibitemNoStop [0]{.\EOS\space}%
\providecommand \EOS [0]{\spacefactor3000\relax}%
\providecommand \BibitemShut  [1]{\csname bibitem#1\endcsname}%
\let\auto@bib@innerbib\@empty
\bibitem [{\citenamefont {Moore}(2018)}]{Moore2018}%
  \BibitemOpen
  \bibfield  {author} {\bibinfo {author} {\bibfnamefont {S.~K.}\ \bibnamefont
  {Moore}},\ }\href {https://doi.org/10.1109/MSPEC.2018.8241736} {\bibfield
  {journal} {\bibinfo  {journal} {IEEE Spectrum}\ }\textbf {\bibinfo {volume}
  {55}},\ \bibinfo {pages} {46} (\bibinfo {year} {2018})}\BibitemShut {NoStop}%
\bibitem [{\citenamefont {Bakshi}(2018)}]{Bakshi2018}%
  \BibitemOpen
  \bibinfo {editor} {\bibfnamefont {V.}~\bibnamefont {Bakshi}},\ ed.,\
  \href@noop {} {\emph {\bibinfo {title} {{EUV Lithography}}}},\ \bibinfo
  {edition} {{2nd Edition}}\ ed.\ (\bibinfo  {publisher} {SPIE Press},\
  \bibinfo {year} {2018})\BibitemShut {NoStop}%
\bibitem [{\citenamefont {Fomenkov}\ \emph {et~al.}(2017)\citenamefont
  {Fomenkov}, \citenamefont {Brandt}, \citenamefont {Ershov}, \citenamefont
  {Schafgans}, \citenamefont {Tao}, \citenamefont {Vaschenko}, \citenamefont
  {Rokitski}, \citenamefont {Kats}, \citenamefont {Vargas}, \citenamefont
  {Purvis}, \citenamefont {Rafac}, \citenamefont {La~Fontaine}, \citenamefont
  {De~Dea}, \citenamefont {LaForge}, \citenamefont {Stewart}, \citenamefont
  {Chang}, \citenamefont {Graham}, \citenamefont {Riggs}, \citenamefont
  {Taylor}, \citenamefont {Abraham},\ and\ \citenamefont
  {Brown}}]{Fomenkov2017}%
  \BibitemOpen
  \bibfield  {author} {\bibinfo {author} {\bibfnamefont {I.}~\bibnamefont
  {Fomenkov}}, \bibinfo {author} {\bibfnamefont {D.}~\bibnamefont {Brandt}},
  \bibinfo {author} {\bibfnamefont {A.}~\bibnamefont {Ershov}}, \bibinfo
  {author} {\bibfnamefont {A.}~\bibnamefont {Schafgans}}, \bibinfo {author}
  {\bibfnamefont {Y.}~\bibnamefont {Tao}}, \bibinfo {author} {\bibfnamefont
  {G.}~\bibnamefont {Vaschenko}}, \bibinfo {author} {\bibfnamefont
  {S.}~\bibnamefont {Rokitski}}, \bibinfo {author} {\bibfnamefont
  {M.}~\bibnamefont {Kats}}, \bibinfo {author} {\bibfnamefont {M.}~\bibnamefont
  {Vargas}}, \bibinfo {author} {\bibfnamefont {M.}~\bibnamefont {Purvis}},
  \bibinfo {author} {\bibfnamefont {R.}~\bibnamefont {Rafac}}, \bibinfo
  {author} {\bibfnamefont {B.}~\bibnamefont {La~Fontaine}}, \bibinfo {author}
  {\bibfnamefont {S.}~\bibnamefont {De~Dea}}, \bibinfo {author} {\bibfnamefont
  {A.}~\bibnamefont {LaForge}}, \bibinfo {author} {\bibfnamefont
  {J.}~\bibnamefont {Stewart}}, \bibinfo {author} {\bibfnamefont
  {S.}~\bibnamefont {Chang}}, \bibinfo {author} {\bibfnamefont
  {M.}~\bibnamefont {Graham}}, \bibinfo {author} {\bibfnamefont
  {D.}~\bibnamefont {Riggs}}, \bibinfo {author} {\bibfnamefont
  {T.}~\bibnamefont {Taylor}}, \bibinfo {author} {\bibfnamefont
  {M.}~\bibnamefont {Abraham}}, \ and\ \bibinfo {author} {\bibfnamefont
  {D.}~\bibnamefont {Brown}},\ }\href {\doibase 10.1515/aot-2017-0029}
  {\bibfield  {journal} {\bibinfo  {journal} {Adv. Opt. Techn.}\ }\textbf
  {\bibinfo {volume} {6}},\ \bibinfo {pages} {173} (\bibinfo {year}
  {2017})}\BibitemShut {NoStop}%
\bibitem [{\citenamefont {Bajt}\ \emph {et~al.}(2002)\citenamefont {Bajt},
  \citenamefont {Alameda}, \citenamefont {Barbee}, \citenamefont {Clift},
  \citenamefont {Folta}, \citenamefont {Kaufmann},\ and\ \citenamefont
  {Spiller}}]{Bajt2002}%
  \BibitemOpen
  \bibfield  {author} {\bibinfo {author} {\bibfnamefont {S.}~\bibnamefont
  {Bajt}}, \bibinfo {author} {\bibfnamefont {J.~B.}\ \bibnamefont {Alameda}},
  \bibinfo {author} {\bibfnamefont {T.~W.}\ \bibnamefont {Barbee},
  \bibfnamefont {Jr.}}, \bibinfo {author} {\bibfnamefont {W.~M.}\ \bibnamefont
  {Clift}}, \bibinfo {author} {\bibfnamefont {J.~A.}\ \bibnamefont {Folta}},
  \bibinfo {author} {\bibfnamefont {B.}~\bibnamefont {Kaufmann}}, \ and\
  \bibinfo {author} {\bibfnamefont {E.~A.}\ \bibnamefont {Spiller}},\ }\href
  {\doibase 10.1117/1.1489426} {\bibfield  {journal} {\bibinfo  {journal} {Opt.
  Eng.}\ }\textbf {\bibinfo {volume} {41}},\ \bibinfo {pages} {1797} (\bibinfo
  {year} {2002})}\BibitemShut {NoStop}%
\bibitem [{\citenamefont {Huang}\ \emph {et~al.}(2017)\citenamefont {Huang},
  \citenamefont {Medvedev}, \citenamefont {van~de Kruijs}, \citenamefont
  {Yakshin}, \citenamefont {Louis},\ and\ \citenamefont {Bijkerk}}]{Huang2017}%
  \BibitemOpen
  \bibfield  {author} {\bibinfo {author} {\bibfnamefont {Q.}~\bibnamefont
  {Huang}}, \bibinfo {author} {\bibfnamefont {V.}~\bibnamefont {Medvedev}},
  \bibinfo {author} {\bibfnamefont {R.}~\bibnamefont {van~de Kruijs}}, \bibinfo
  {author} {\bibfnamefont {A.}~\bibnamefont {Yakshin}}, \bibinfo {author}
  {\bibfnamefont {E.}~\bibnamefont {Louis}}, \ and\ \bibinfo {author}
  {\bibfnamefont {F.}~\bibnamefont {Bijkerk}},\ }\href {\doibase
  10.1063/1.4978290} {\bibfield  {journal} {\bibinfo  {journal} {Appl. Phys.
  Rev.}\ }\textbf {\bibinfo {volume} {4}},\ \bibinfo {pages} {011104} (\bibinfo
  {year} {2017})}\BibitemShut {NoStop}%
\bibitem [{\citenamefont {Fujioka}\ \emph {et~al.}(2008)\citenamefont
  {Fujioka}, \citenamefont {Shimomura}, \citenamefont {Shimada}, \citenamefont
  {Maeda}, \citenamefont {Sakaguchi}, \citenamefont {Nakai}, \citenamefont
  {Aota}, \citenamefont {Nishimura}, \citenamefont {Ozaki}, \citenamefont
  {Sunahara}, \citenamefont {Nishihara}, \citenamefont {Miyanaga},
  \citenamefont {Izawa},\ and\ \citenamefont {Mima}}]{Fujioka2008}%
  \BibitemOpen
  \bibfield  {author} {\bibinfo {author} {\bibfnamefont {S.}~\bibnamefont
  {Fujioka}}, \bibinfo {author} {\bibfnamefont {M.}~\bibnamefont {Shimomura}},
  \bibinfo {author} {\bibfnamefont {Y.}~\bibnamefont {Shimada}}, \bibinfo
  {author} {\bibfnamefont {S.}~\bibnamefont {Maeda}}, \bibinfo {author}
  {\bibfnamefont {H.}~\bibnamefont {Sakaguchi}}, \bibinfo {author}
  {\bibfnamefont {Y.}~\bibnamefont {Nakai}}, \bibinfo {author} {\bibfnamefont
  {T.}~\bibnamefont {Aota}}, \bibinfo {author} {\bibfnamefont {H.}~\bibnamefont
  {Nishimura}}, \bibinfo {author} {\bibfnamefont {N.}~\bibnamefont {Ozaki}},
  \bibinfo {author} {\bibfnamefont {A.}~\bibnamefont {Sunahara}}, \bibinfo
  {author} {\bibfnamefont {K.}~\bibnamefont {Nishihara}}, \bibinfo {author}
  {\bibfnamefont {N.}~\bibnamefont {Miyanaga}}, \bibinfo {author}
  {\bibfnamefont {Y.}~\bibnamefont {Izawa}}, \ and\ \bibinfo {author}
  {\bibfnamefont {K.}~\bibnamefont {Mima}},\ }\href {\doibase
  10.1063/1.2948874} {\bibfield  {journal} {\bibinfo  {journal} {Appl. Phys.
  Lett.}\ }\textbf {\bibinfo {volume} {92}},\ \bibinfo {pages} {241502}
  (\bibinfo {year} {2008})}\BibitemShut {NoStop}%
\bibitem [{\citenamefont {Nishihara}\ \emph {et~al.}(2008)\citenamefont
  {Nishihara}, \citenamefont {Sunahara}, \citenamefont {Sasaki}, \citenamefont
  {Nunami}, \citenamefont {Tanuma}, \citenamefont {Fujioka}, \citenamefont
  {Shimada}, \citenamefont {Fujima}, \citenamefont {Furukawa}, \citenamefont
  {Kato}, \citenamefont {Koike}, \citenamefont {More}, \citenamefont
  {Murakami}, \citenamefont {Nishikawa}, \citenamefont {Zhakhovskii},
  \citenamefont {Gamata}, \citenamefont {Takata}, \citenamefont {Ueda},
  \citenamefont {Nishimura}, \citenamefont {Izawa}, \citenamefont {Miyanaga},\
  and\ \citenamefont {Mima}}]{Nishihara2008}%
  \BibitemOpen
  \bibfield  {author} {\bibinfo {author} {\bibfnamefont {K.}~\bibnamefont
  {Nishihara}}, \bibinfo {author} {\bibfnamefont {A.}~\bibnamefont {Sunahara}},
  \bibinfo {author} {\bibfnamefont {A.}~\bibnamefont {Sasaki}}, \bibinfo
  {author} {\bibfnamefont {M.}~\bibnamefont {Nunami}}, \bibinfo {author}
  {\bibfnamefont {H.}~\bibnamefont {Tanuma}}, \bibinfo {author} {\bibfnamefont
  {S.}~\bibnamefont {Fujioka}}, \bibinfo {author} {\bibfnamefont
  {Y.}~\bibnamefont {Shimada}}, \bibinfo {author} {\bibfnamefont
  {K.}~\bibnamefont {Fujima}}, \bibinfo {author} {\bibfnamefont
  {H.}~\bibnamefont {Furukawa}}, \bibinfo {author} {\bibfnamefont
  {T.}~\bibnamefont {Kato}}, \bibinfo {author} {\bibfnamefont {F.}~\bibnamefont
  {Koike}}, \bibinfo {author} {\bibfnamefont {R.}~\bibnamefont {More}},
  \bibinfo {author} {\bibfnamefont {M.}~\bibnamefont {Murakami}}, \bibinfo
  {author} {\bibfnamefont {T.}~\bibnamefont {Nishikawa}}, \bibinfo {author}
  {\bibfnamefont {V.}~\bibnamefont {Zhakhovskii}}, \bibinfo {author}
  {\bibfnamefont {K.}~\bibnamefont {Gamata}}, \bibinfo {author} {\bibfnamefont
  {A.}~\bibnamefont {Takata}}, \bibinfo {author} {\bibfnamefont
  {H.}~\bibnamefont {Ueda}}, \bibinfo {author} {\bibfnamefont {H.}~\bibnamefont
  {Nishimura}}, \bibinfo {author} {\bibfnamefont {Y.}~\bibnamefont {Izawa}},
  \bibinfo {author} {\bibfnamefont {N.}~\bibnamefont {Miyanaga}}, \ and\
  \bibinfo {author} {\bibfnamefont {K.}~\bibnamefont {Mima}},\ }\href {\doibase
  10.1063/1.2907154} {\bibfield  {journal} {\bibinfo  {journal} {Phys.
  Plasmas}\ }\textbf {\bibinfo {volume} {15}},\ \bibinfo {pages} {056708}
  (\bibinfo {year} {2008})}\BibitemShut {NoStop}%
\bibitem [{\citenamefont {Fujimoto}\ \emph {et~al.}(2012)\citenamefont
  {Fujimoto}, \citenamefont {Mizoguchi}, \citenamefont {Abe}, \citenamefont
  {Tanaka}, \citenamefont {Ohta}, \citenamefont {Hori}, \citenamefont
  {Yanagida},\ and\ \citenamefont {Nakarai}}]{Fujimoto2012}%
  \BibitemOpen
  \bibfield  {author} {\bibinfo {author} {\bibfnamefont {J.}~\bibnamefont
  {Fujimoto}}, \bibinfo {author} {\bibfnamefont {H.}~\bibnamefont {Mizoguchi}},
  \bibinfo {author} {\bibfnamefont {T.}~\bibnamefont {Abe}}, \bibinfo {author}
  {\bibfnamefont {S.}~\bibnamefont {Tanaka}}, \bibinfo {author} {\bibfnamefont
  {T.}~\bibnamefont {Ohta}}, \bibinfo {author} {\bibfnamefont {T.}~\bibnamefont
  {Hori}}, \bibinfo {author} {\bibfnamefont {T.}~\bibnamefont {Yanagida}}, \
  and\ \bibinfo {author} {\bibfnamefont {H.}~\bibnamefont {Nakarai}},\ }\href
  {\doibase 10.1117/1.JMM.11.2.021111} {\bibfield  {journal} {\bibinfo
  {journal} {Journal of Micro/Nanolithography, MEMS, and MOEMS}\ }\textbf
  {\bibinfo {volume} {11}},\ \bibinfo {pages} {1 } (\bibinfo {year}
  {2012})}\BibitemShut {NoStop}%
\bibitem [{\citenamefont {Kurilovich}\ \emph {et~al.}(2016)\citenamefont
  {Kurilovich}, \citenamefont {Klein}, \citenamefont {Torretti}, \citenamefont
  {Lassise}, \citenamefont {Hoekstra}, \citenamefont {Ubachs}, \citenamefont
  {Gelderblom},\ and\ \citenamefont {Versolato}}]{Kurilovich2016}%
  \BibitemOpen
  \bibfield  {author} {\bibinfo {author} {\bibfnamefont {D.}~\bibnamefont
  {Kurilovich}}, \bibinfo {author} {\bibfnamefont {A.~L.}\ \bibnamefont
  {Klein}}, \bibinfo {author} {\bibfnamefont {F.}~\bibnamefont {Torretti}},
  \bibinfo {author} {\bibfnamefont {A.}~\bibnamefont {Lassise}}, \bibinfo
  {author} {\bibfnamefont {R.}~\bibnamefont {Hoekstra}}, \bibinfo {author}
  {\bibfnamefont {W.}~\bibnamefont {Ubachs}}, \bibinfo {author} {\bibfnamefont
  {H.}~\bibnamefont {Gelderblom}}, \ and\ \bibinfo {author} {\bibfnamefont
  {O.~O.}\ \bibnamefont {Versolato}},\ }\href {\doibase
  10.1103/PhysRevApplied.6.014018} {\bibfield  {journal} {\bibinfo  {journal}
  {Phys. Rev. Appl.}\ }\textbf {\bibinfo {volume} {6}},\ \bibinfo {pages}
  {014018} (\bibinfo {year} {2016})}\BibitemShut {NoStop}%
\bibitem [{\citenamefont {Grigoryev}\ \emph {et~al.}(2018)\citenamefont
  {Grigoryev}, \citenamefont {Lakatosh}, \citenamefont {Krivokorytov},
  \citenamefont {Zhakhovsky}, \citenamefont {Dyachkov}, \citenamefont
  {Ilnitsky}, \citenamefont {Migdal}, \citenamefont {Inogamov}, \citenamefont
  {Vinokhodov}, \citenamefont {Kompanets} \emph {et~al.}}]{Grigoryev2018}%
  \BibitemOpen
  \bibfield  {author} {\bibinfo {author} {\bibfnamefont {S.~Y.}\ \bibnamefont
  {Grigoryev}}, \bibinfo {author} {\bibfnamefont {B.}~\bibnamefont {Lakatosh}},
  \bibinfo {author} {\bibfnamefont {M.}~\bibnamefont {Krivokorytov}}, \bibinfo
  {author} {\bibfnamefont {V.}~\bibnamefont {Zhakhovsky}}, \bibinfo {author}
  {\bibfnamefont {S.}~\bibnamefont {Dyachkov}}, \bibinfo {author}
  {\bibfnamefont {D.}~\bibnamefont {Ilnitsky}}, \bibinfo {author}
  {\bibfnamefont {K.}~\bibnamefont {Migdal}}, \bibinfo {author} {\bibfnamefont
  {N.}~\bibnamefont {Inogamov}}, \bibinfo {author} {\bibfnamefont {A.~Y.}\
  \bibnamefont {Vinokhodov}}, \bibinfo {author} {\bibfnamefont
  {V.}~\bibnamefont {Kompanets}},  \emph {et~al.},\ }\href@noop {} {\bibfield
  {journal} {\bibinfo  {journal} {Phys. Rev. Appl.}\ }\textbf {\bibinfo
  {volume} {10}},\ \bibinfo {pages} {064009} (\bibinfo {year}
  {2018})}\BibitemShut {NoStop}%
\bibitem [{\citenamefont {Hernandez-Rueda}\ \emph {et~al.}(2022)\citenamefont
  {Hernandez-Rueda}, \citenamefont {Liu}, \citenamefont {Hemminga},
  \citenamefont {Mostafa}, \citenamefont {Meijer}, \citenamefont {Kurilovich},
  \citenamefont {Basko}, \citenamefont {Gelderblom}, \citenamefont {Sheil},\
  and\ \citenamefont {Versolato}}]{HernandezRueda2022}%
  \BibitemOpen
  \bibfield  {author} {\bibinfo {author} {\bibfnamefont {J.}~\bibnamefont
  {Hernandez-Rueda}}, \bibinfo {author} {\bibfnamefont {B.}~\bibnamefont
  {Liu}}, \bibinfo {author} {\bibfnamefont {D.~J.}\ \bibnamefont {Hemminga}},
  \bibinfo {author} {\bibfnamefont {Y.}~\bibnamefont {Mostafa}}, \bibinfo
  {author} {\bibfnamefont {R.~A.}\ \bibnamefont {Meijer}}, \bibinfo {author}
  {\bibfnamefont {D.}~\bibnamefont {Kurilovich}}, \bibinfo {author}
  {\bibfnamefont {M.}~\bibnamefont {Basko}}, \bibinfo {author} {\bibfnamefont
  {H.}~\bibnamefont {Gelderblom}}, \bibinfo {author} {\bibfnamefont
  {J.}~\bibnamefont {Sheil}}, \ and\ \bibinfo {author} {\bibfnamefont {O.~O.}\
  \bibnamefont {Versolato}},\ }\href {\doibase
  10.1103/PhysRevResearch.4.013142} {\bibfield  {journal} {\bibinfo  {journal}
  {Phys. Rev. Research}\ }\textbf {\bibinfo {volume} {4}},\ \bibinfo {pages}
  {013142} (\bibinfo {year} {2022})}\BibitemShut {NoStop}%
\bibitem [{\citenamefont {Versolato}(2019)}]{Versolato2019}%
  \BibitemOpen
  \bibfield  {author} {\bibinfo {author} {\bibfnamefont {O.~O.}\ \bibnamefont
  {Versolato}},\ }\href {\doibase 10.1088/1361-6595/ab3302} {\bibfield
  {journal} {\bibinfo  {journal} {Plasma Sources Sci. Tech.}\ }\textbf
  {\bibinfo {volume} {28}},\ \bibinfo {pages} {083001} (\bibinfo {year}
  {2019})}\BibitemShut {NoStop}%
\bibitem [{\citenamefont {Svendsen}\ and\ \citenamefont
  {O'Sullivan}(1994)}]{Svendsen1994}%
  \BibitemOpen
  \bibfield  {author} {\bibinfo {author} {\bibfnamefont {W.}~\bibnamefont
  {Svendsen}}\ and\ \bibinfo {author} {\bibfnamefont {G.}~\bibnamefont
  {O'Sullivan}},\ }\href {\doibase 10.1103/PhysRevA.50.3710} {\bibfield
  {journal} {\bibinfo  {journal} {Phys. Rev. A}\ }\textbf {\bibinfo {volume}
  {50}},\ \bibinfo {pages} {3710} (\bibinfo {year} {1994})}\BibitemShut
  {NoStop}%
\bibitem [{\citenamefont {Churilov}\ and\ \citenamefont
  {Ryabtsev}(2006)}]{Churilov2006b}%
  \BibitemOpen
  \bibfield  {author} {\bibinfo {author} {\bibfnamefont {S.~S.}\ \bibnamefont
  {Churilov}}\ and\ \bibinfo {author} {\bibfnamefont {A.~N.}\ \bibnamefont
  {Ryabtsev}},\ }\href {\doibase 10.1088/0031-8949/73/6/014} {\bibfield
  {journal} {\bibinfo  {journal} {Phys. Scr.}\ }\textbf {\bibinfo {volume}
  {73}},\ \bibinfo {pages} {614} (\bibinfo {year} {2006})}\BibitemShut
  {NoStop}%
\bibitem [{\citenamefont {Fujioka}\ \emph
  {et~al.}(2005{\natexlab{a}})\citenamefont {Fujioka}, \citenamefont
  {Nishimura}, \citenamefont {Nishihara}, \citenamefont {Sasaki}, \citenamefont
  {Sunahara}, \citenamefont {Okuno}, \citenamefont {Ueda}, \citenamefont
  {Ando}, \citenamefont {Tao}, \citenamefont {Shimada}, \citenamefont
  {Hashimoto}, \citenamefont {Yamaura}, \citenamefont {Shigemori},
  \citenamefont {Nakai}, \citenamefont {Nagai}, \citenamefont {Norimatsu},
  \citenamefont {Nishikawa}, \citenamefont {Miyanaga}, \citenamefont {Izawa},\
  and\ \citenamefont {Mima}}]{Fujioka2005a}%
  \BibitemOpen
  \bibfield  {author} {\bibinfo {author} {\bibfnamefont {S.}~\bibnamefont
  {Fujioka}}, \bibinfo {author} {\bibfnamefont {H.}~\bibnamefont {Nishimura}},
  \bibinfo {author} {\bibfnamefont {K.}~\bibnamefont {Nishihara}}, \bibinfo
  {author} {\bibfnamefont {A.}~\bibnamefont {Sasaki}}, \bibinfo {author}
  {\bibfnamefont {A.}~\bibnamefont {Sunahara}}, \bibinfo {author}
  {\bibfnamefont {T.}~\bibnamefont {Okuno}}, \bibinfo {author} {\bibfnamefont
  {N.}~\bibnamefont {Ueda}}, \bibinfo {author} {\bibfnamefont {T.}~\bibnamefont
  {Ando}}, \bibinfo {author} {\bibfnamefont {Y.}~\bibnamefont {Tao}}, \bibinfo
  {author} {\bibfnamefont {Y.}~\bibnamefont {Shimada}}, \bibinfo {author}
  {\bibfnamefont {K.}~\bibnamefont {Hashimoto}}, \bibinfo {author}
  {\bibfnamefont {M.}~\bibnamefont {Yamaura}}, \bibinfo {author} {\bibfnamefont
  {K.}~\bibnamefont {Shigemori}}, \bibinfo {author} {\bibfnamefont
  {M.}~\bibnamefont {Nakai}}, \bibinfo {author} {\bibfnamefont
  {K.}~\bibnamefont {Nagai}}, \bibinfo {author} {\bibfnamefont
  {T.}~\bibnamefont {Norimatsu}}, \bibinfo {author} {\bibfnamefont
  {T.}~\bibnamefont {Nishikawa}}, \bibinfo {author} {\bibfnamefont
  {N.}~\bibnamefont {Miyanaga}}, \bibinfo {author} {\bibfnamefont
  {Y.}~\bibnamefont {Izawa}}, \ and\ \bibinfo {author} {\bibfnamefont
  {K.}~\bibnamefont {Mima}},\ }\href {\doibase 10.1103/PhysRevLett.95.235004}
  {\bibfield  {journal} {\bibinfo  {journal} {Phys. Rev. Lett.}\ }\textbf
  {\bibinfo {volume} {95}},\ \bibinfo {pages} {235004} (\bibinfo {year}
  {2005}{\natexlab{a}})}\BibitemShut {NoStop}%
\bibitem [{\citenamefont {Sasaki}\ \emph {et~al.}(2010)\citenamefont {Sasaki},
  \citenamefont {Sunahara}, \citenamefont {Furukawa}, \citenamefont
  {Nishihara}, \citenamefont {Fujioka}, \citenamefont {Nishikawa},
  \citenamefont {Koike}, \citenamefont {Ohashi},\ and\ \citenamefont
  {Tanuma}}]{Sasaki2010}%
  \BibitemOpen
  \bibfield  {author} {\bibinfo {author} {\bibfnamefont {A.}~\bibnamefont
  {Sasaki}}, \bibinfo {author} {\bibfnamefont {A.}~\bibnamefont {Sunahara}},
  \bibinfo {author} {\bibfnamefont {H.}~\bibnamefont {Furukawa}}, \bibinfo
  {author} {\bibfnamefont {K.}~\bibnamefont {Nishihara}}, \bibinfo {author}
  {\bibfnamefont {S.}~\bibnamefont {Fujioka}}, \bibinfo {author} {\bibfnamefont
  {T.}~\bibnamefont {Nishikawa}}, \bibinfo {author} {\bibfnamefont
  {F.}~\bibnamefont {Koike}}, \bibinfo {author} {\bibfnamefont
  {H.}~\bibnamefont {Ohashi}}, \ and\ \bibinfo {author} {\bibfnamefont
  {H.}~\bibnamefont {Tanuma}},\ }\href {\doibase 10.1063/1.3373427} {\bibfield
  {journal} {\bibinfo  {journal} {J. Appl. Phys.}\ }\textbf {\bibinfo {volume}
  {107}},\ \bibinfo {pages} {113303} (\bibinfo {year} {2010})}\BibitemShut
  {NoStop}%
\bibitem [{\citenamefont {O'Sullivan}\ \emph {et~al.}(2015)\citenamefont
  {O'Sullivan}, \citenamefont {Li}, \citenamefont {D'Arcy}, \citenamefont
  {Dunne}, \citenamefont {Hayden}, \citenamefont {Kilbane}, \citenamefont
  {McCormack}, \citenamefont {Ohashi}, \citenamefont {O'Reilly}, \citenamefont
  {Sheridan}, \citenamefont {Sokell}, \citenamefont {Suzuki},\ and\
  \citenamefont {Higashiguchi}}]{OSullivan2015}%
  \BibitemOpen
  \bibfield  {author} {\bibinfo {author} {\bibfnamefont {G.}~\bibnamefont
  {O'Sullivan}}, \bibinfo {author} {\bibfnamefont {B.}~\bibnamefont {Li}},
  \bibinfo {author} {\bibfnamefont {R.}~\bibnamefont {D'Arcy}}, \bibinfo
  {author} {\bibfnamefont {P.}~\bibnamefont {Dunne}}, \bibinfo {author}
  {\bibfnamefont {P.}~\bibnamefont {Hayden}}, \bibinfo {author} {\bibfnamefont
  {D.}~\bibnamefont {Kilbane}}, \bibinfo {author} {\bibfnamefont
  {T.}~\bibnamefont {McCormack}}, \bibinfo {author} {\bibfnamefont
  {H.}~\bibnamefont {Ohashi}}, \bibinfo {author} {\bibfnamefont
  {F.}~\bibnamefont {O'Reilly}}, \bibinfo {author} {\bibfnamefont
  {P.}~\bibnamefont {Sheridan}}, \bibinfo {author} {\bibfnamefont
  {E.}~\bibnamefont {Sokell}}, \bibinfo {author} {\bibfnamefont
  {C.}~\bibnamefont {Suzuki}}, \ and\ \bibinfo {author} {\bibfnamefont
  {T.}~\bibnamefont {Higashiguchi}},\ }\href {\doibase
  10.1088/0953-4075/48/14/144025} {\bibfield  {journal} {\bibinfo  {journal}
  {J. Phys. B: At. Mol. Opt. Phys.}\ }\textbf {\bibinfo {volume} {48}},\
  \bibinfo {pages} {144025} (\bibinfo {year} {2015})}\BibitemShut {NoStop}%
\bibitem [{\citenamefont {Scheers}\ \emph {et~al.}(2020)\citenamefont
  {Scheers}, \citenamefont {Shah}, \citenamefont {Ryabtsev}, \citenamefont
  {Bekker}, \citenamefont {Torretti}, \citenamefont {Sheil}, \citenamefont
  {Czapski}, \citenamefont {Berengut}, \citenamefont {Ubachs}, \citenamefont
  {L\'opez-Urrutia}, \citenamefont {Hoekstra},\ and\ \citenamefont
  {Versolato}}]{Scheers2020}%
  \BibitemOpen
  \bibfield  {author} {\bibinfo {author} {\bibfnamefont {J.}~\bibnamefont
  {Scheers}}, \bibinfo {author} {\bibfnamefont {C.}~\bibnamefont {Shah}},
  \bibinfo {author} {\bibfnamefont {A.}~\bibnamefont {Ryabtsev}}, \bibinfo
  {author} {\bibfnamefont {H.}~\bibnamefont {Bekker}}, \bibinfo {author}
  {\bibfnamefont {F.}~\bibnamefont {Torretti}}, \bibinfo {author}
  {\bibfnamefont {J.}~\bibnamefont {Sheil}}, \bibinfo {author} {\bibfnamefont
  {D.~A.}\ \bibnamefont {Czapski}}, \bibinfo {author} {\bibfnamefont {J.~C.}\
  \bibnamefont {Berengut}}, \bibinfo {author} {\bibfnamefont {W.}~\bibnamefont
  {Ubachs}}, \bibinfo {author} {\bibfnamefont {J.~R.~C.}\ \bibnamefont
  {L\'opez-Urrutia}}, \bibinfo {author} {\bibfnamefont {R.}~\bibnamefont
  {Hoekstra}}, \ and\ \bibinfo {author} {\bibfnamefont {O.~O.}\ \bibnamefont
  {Versolato}},\ }\href {\doibase 10.1103/PhysRevA.101.062511} {\bibfield
  {journal} {\bibinfo  {journal} {Phys. Rev. A}\ }\textbf {\bibinfo {volume}
  {101}},\ \bibinfo {pages} {062511} (\bibinfo {year} {2020})}\BibitemShut
  {NoStop}%
\bibitem [{\citenamefont {Torretti}\ \emph {et~al.}(2020)\citenamefont
  {Torretti}, \citenamefont {Sheil}, \citenamefont {Schupp}, \citenamefont
  {Basko}, \citenamefont {Bayraktar}, \citenamefont {Meijer}, \citenamefont
  {Witte}, \citenamefont {Ubachs}, \citenamefont {Hoekstra}, \citenamefont
  {Versolato}, \citenamefont {Neukirch},\ and\ \citenamefont
  {Colgan}}]{Torretti2020}%
  \BibitemOpen
  \bibfield  {author} {\bibinfo {author} {\bibfnamefont {F.}~\bibnamefont
  {Torretti}}, \bibinfo {author} {\bibfnamefont {J.}~\bibnamefont {Sheil}},
  \bibinfo {author} {\bibfnamefont {R.}~\bibnamefont {Schupp}}, \bibinfo
  {author} {\bibfnamefont {M.~M.}\ \bibnamefont {Basko}}, \bibinfo {author}
  {\bibfnamefont {M.}~\bibnamefont {Bayraktar}}, \bibinfo {author}
  {\bibfnamefont {R.~A.}\ \bibnamefont {Meijer}}, \bibinfo {author}
  {\bibfnamefont {S.}~\bibnamefont {Witte}}, \bibinfo {author} {\bibfnamefont
  {W.}~\bibnamefont {Ubachs}}, \bibinfo {author} {\bibfnamefont
  {R.}~\bibnamefont {Hoekstra}}, \bibinfo {author} {\bibfnamefont {O.~O.}\
  \bibnamefont {Versolato}}, \bibinfo {author} {\bibfnamefont {A.~J.}\
  \bibnamefont {Neukirch}}, \ and\ \bibinfo {author} {\bibfnamefont
  {J.}~\bibnamefont {Colgan}},\ }\href {\doibase 10.1038/s41467-020-15678-y}
  {\bibfield  {journal} {\bibinfo  {journal} {Nat. Commun.}\ }\textbf {\bibinfo
  {volume} {11}} (\bibinfo {year} {2020}),\
  10.1038/s41467-020-15678-y}\BibitemShut {NoStop}%
\bibitem [{\citenamefont {Fomenkov}\ \emph {et~al.}(2018)\citenamefont
  {Fomenkov}, \citenamefont {Purvis}, \citenamefont {Schafgans}, \citenamefont
  {Tao}, \citenamefont {Rokitski}, \citenamefont {Stewart}, \citenamefont
  {LaForge}, \citenamefont {Ershov}, \citenamefont {Rafac}, \citenamefont
  {Dea}, \citenamefont {Rajyaguru}, \citenamefont {Vaschenko}, \citenamefont
  {Abraham}, \citenamefont {Brandt},\ and\ \citenamefont
  {Brown}}]{Fomenkov2018}%
  \BibitemOpen
  \bibfield  {author} {\bibinfo {author} {\bibfnamefont {I.~V.}\ \bibnamefont
  {Fomenkov}}, \bibinfo {author} {\bibfnamefont {M.~A.}\ \bibnamefont
  {Purvis}}, \bibinfo {author} {\bibfnamefont {A.~A.}\ \bibnamefont
  {Schafgans}}, \bibinfo {author} {\bibfnamefont {Y.}~\bibnamefont {Tao}},
  \bibinfo {author} {\bibfnamefont {S.}~\bibnamefont {Rokitski}}, \bibinfo
  {author} {\bibfnamefont {J.}~\bibnamefont {Stewart}}, \bibinfo {author}
  {\bibfnamefont {A.}~\bibnamefont {LaForge}}, \bibinfo {author} {\bibfnamefont
  {A.~I.}\ \bibnamefont {Ershov}}, \bibinfo {author} {\bibfnamefont {R.~J.}\
  \bibnamefont {Rafac}}, \bibinfo {author} {\bibfnamefont {S.~D.}\ \bibnamefont
  {Dea}}, \bibinfo {author} {\bibfnamefont {C.}~\bibnamefont {Rajyaguru}},
  \bibinfo {author} {\bibfnamefont {G.~O.}\ \bibnamefont {Vaschenko}}, \bibinfo
  {author} {\bibfnamefont {M.}~\bibnamefont {Abraham}}, \bibinfo {author}
  {\bibfnamefont {D.~C.}\ \bibnamefont {Brandt}}, \ and\ \bibinfo {author}
  {\bibfnamefont {D.~J.}\ \bibnamefont {Brown}},\ }in\ \href {\doibase
  10.1117/12.2502801} {\emph {\bibinfo {booktitle} {International Conference on
  Extreme Ultraviolet Lithography 2018}}},\ Vol.\ \bibinfo {volume} {10809},\
  \bibinfo {editor} {edited by\ \bibinfo {editor} {\bibfnamefont {K.~G.}\
  \bibnamefont {Ronse}}, \bibinfo {editor} {\bibfnamefont {E.}~\bibnamefont
  {Hendrickx}}, \bibinfo {editor} {\bibfnamefont {P.~P.}\ \bibnamefont
  {Naulleau}}, \bibinfo {editor} {\bibfnamefont {P.~A.}\ \bibnamefont
  {Gargini}}, \ and\ \bibinfo {editor} {\bibfnamefont {T.}~\bibnamefont
  {Itani}}},\ \bibinfo {organization} {International Society for Optics and
  Photonics}\ (\bibinfo  {publisher} {SPIE},\ \bibinfo {year} {2018})\ pp.\
  \bibinfo {pages} {213 -- 213}\BibitemShut {NoStop}%
\bibitem [{\citenamefont {Brandt}\ \emph {et~al.}(2021)\citenamefont {Brandt},
  \citenamefont {Purvis}, \citenamefont {Fomenkov}, \citenamefont {Brown},
  \citenamefont {Schafgans}, \citenamefont {Mayer},\ and\ \citenamefont
  {Rafac}}]{Brandt2022}%
  \BibitemOpen
  \bibfield  {author} {\bibinfo {author} {\bibfnamefont {D.~C.}\ \bibnamefont
  {Brandt}}, \bibinfo {author} {\bibfnamefont {M.}~\bibnamefont {Purvis}},
  \bibinfo {author} {\bibfnamefont {I.}~\bibnamefont {Fomenkov}}, \bibinfo
  {author} {\bibfnamefont {D.}~\bibnamefont {Brown}}, \bibinfo {author}
  {\bibfnamefont {A.}~\bibnamefont {Schafgans}}, \bibinfo {author}
  {\bibfnamefont {P.}~\bibnamefont {Mayer}}, \ and\ \bibinfo {author}
  {\bibfnamefont {R.}~\bibnamefont {Rafac}},\ }in\ \href {\doibase
  10.1117/12.2584413} {\emph {\bibinfo {booktitle} {Extreme Ultraviolet (EUV)
  Lithography XII}}},\ Vol.\ \bibinfo {volume} {11609},\ \bibinfo {editor}
  {edited by\ \bibinfo {editor} {\bibfnamefont {N.~M.}\ \bibnamefont {Felix}}\
  and\ \bibinfo {editor} {\bibfnamefont {A.}~\bibnamefont {Lio}}},\ \bibinfo
  {organization} {International Society for Optics and Photonics}\ (\bibinfo
  {publisher} {SPIE},\ \bibinfo {year} {2021})\BibitemShut {NoStop}%
\bibitem [{\citenamefont {Nowak}\ \emph {et~al.}(2013)\citenamefont {Nowak},
  \citenamefont {Ohta}, \citenamefont {Suganuma}, \citenamefont {Fujimoto},
  \citenamefont {Mizoguchi}, \citenamefont {Sumitani},\ and\ \citenamefont
  {Endo}}]{Nowak2013}%
  \BibitemOpen
  \bibfield  {author} {\bibinfo {author} {\bibfnamefont {K.}~\bibnamefont
  {Nowak}}, \bibinfo {author} {\bibfnamefont {T.}~\bibnamefont {Ohta}},
  \bibinfo {author} {\bibfnamefont {T.}~\bibnamefont {Suganuma}}, \bibinfo
  {author} {\bibfnamefont {J.}~\bibnamefont {Fujimoto}}, \bibinfo {author}
  {\bibfnamefont {H.}~\bibnamefont {Mizoguchi}}, \bibinfo {author}
  {\bibfnamefont {A.}~\bibnamefont {Sumitani}}, \ and\ \bibinfo {author}
  {\bibfnamefont {A.}~\bibnamefont {Endo}},\ }\href {\doibase
  doi:10.2478/s11772-013-0109-3} {\bibfield  {journal} {\bibinfo  {journal}
  {Opto-Electronics Review}\ }\textbf {\bibinfo {volume} {21}},\ \bibinfo
  {pages} {345} (\bibinfo {year} {2013})}\BibitemShut {NoStop}%
\bibitem [{\citenamefont {Tamer}\ \emph {et~al.}(2021)\citenamefont {Tamer},
  \citenamefont {Reagan}, \citenamefont {Galvin}, \citenamefont {Galbraith},
  \citenamefont {Sistrunk}, \citenamefont {Church}, \citenamefont {Huete},
  \citenamefont {Neurath},\ and\ \citenamefont {Spinka}}]{Tamer2021}%
  \BibitemOpen
  \bibfield  {author} {\bibinfo {author} {\bibfnamefont {I.}~\bibnamefont
  {Tamer}}, \bibinfo {author} {\bibfnamefont {B.~A.}\ \bibnamefont {Reagan}},
  \bibinfo {author} {\bibfnamefont {T.}~\bibnamefont {Galvin}}, \bibinfo
  {author} {\bibfnamefont {J.}~\bibnamefont {Galbraith}}, \bibinfo {author}
  {\bibfnamefont {E.}~\bibnamefont {Sistrunk}}, \bibinfo {author}
  {\bibfnamefont {A.}~\bibnamefont {Church}}, \bibinfo {author} {\bibfnamefont
  {G.}~\bibnamefont {Huete}}, \bibinfo {author} {\bibfnamefont
  {H.}~\bibnamefont {Neurath}}, \ and\ \bibinfo {author} {\bibfnamefont
  {T.}~\bibnamefont {Spinka}},\ }\href {\doibase 10.1364/OL.439238} {\bibfield
  {journal} {\bibinfo  {journal} {Opt. Lett.}\ }\textbf {\bibinfo {volume}
  {46}},\ \bibinfo {pages} {5096} (\bibinfo {year} {2021})}\BibitemShut
  {NoStop}%
\bibitem [{\citenamefont {Sizyuk}\ and\ \citenamefont
  {Hassanein}(2020)}]{Sizyuk2020}%
  \BibitemOpen
  \bibfield  {author} {\bibinfo {author} {\bibfnamefont {T.}~\bibnamefont
  {Sizyuk}}\ and\ \bibinfo {author} {\bibfnamefont {A.}~\bibnamefont
  {Hassanein}},\ }\href {\doibase 10.1063/5.0018576} {\bibfield  {journal}
  {\bibinfo  {journal} {Phys. Plasmas}\ }\textbf {\bibinfo {volume} {27}},\
  \bibinfo {pages} {103507} (\bibinfo {year} {2020})}\BibitemShut {NoStop}%
\bibitem [{\citenamefont {Schupp}\ \emph
  {et~al.}(2021{\natexlab{a}})\citenamefont {Schupp}, \citenamefont {Behnke},
  \citenamefont {Sheil}, \citenamefont {Bouza}, \citenamefont {Bayraktar},
  \citenamefont {Ubachs}, \citenamefont {Hoekstra},\ and\ \citenamefont
  {Versolato}}]{Schupp2021}%
  \BibitemOpen
  \bibfield  {author} {\bibinfo {author} {\bibfnamefont {R.}~\bibnamefont
  {Schupp}}, \bibinfo {author} {\bibfnamefont {L.}~\bibnamefont {Behnke}},
  \bibinfo {author} {\bibfnamefont {J.}~\bibnamefont {Sheil}}, \bibinfo
  {author} {\bibfnamefont {Z.}~\bibnamefont {Bouza}}, \bibinfo {author}
  {\bibfnamefont {M.}~\bibnamefont {Bayraktar}}, \bibinfo {author}
  {\bibfnamefont {W.}~\bibnamefont {Ubachs}}, \bibinfo {author} {\bibfnamefont
  {R.}~\bibnamefont {Hoekstra}}, \ and\ \bibinfo {author} {\bibfnamefont
  {O.~O.}\ \bibnamefont {Versolato}},\ }\href {\doibase
  10.1103/PhysRevResearch.3.013294} {\bibfield  {journal} {\bibinfo  {journal}
  {Phys. Rev. Research}\ }\textbf {\bibinfo {volume} {3}},\ \bibinfo {pages}
  {013294} (\bibinfo {year} {2021}{\natexlab{a}})}\BibitemShut {NoStop}%
\bibitem [{\citenamefont {Yuan}\ \emph {et~al.}(2021)\citenamefont {Yuan},
  \citenamefont {Ma}, \citenamefont {Wang}, \citenamefont {Chen}, \citenamefont
  {Cui}, \citenamefont {Zi}, \citenamefont {Yang}, \citenamefont {Zhang},\ and\
  \citenamefont {Leng}}]{Yuan2021}%
  \BibitemOpen
  \bibfield  {author} {\bibinfo {author} {\bibfnamefont {Y.}~\bibnamefont
  {Yuan}}, \bibinfo {author} {\bibfnamefont {Y.~Y.}\ \bibnamefont {Ma}},
  \bibinfo {author} {\bibfnamefont {W.~P.}\ \bibnamefont {Wang}}, \bibinfo
  {author} {\bibfnamefont {S.~J.}\ \bibnamefont {Chen}}, \bibinfo {author}
  {\bibfnamefont {Y.}~\bibnamefont {Cui}}, \bibinfo {author} {\bibfnamefont
  {M.}~\bibnamefont {Zi}}, \bibinfo {author} {\bibfnamefont {X.~H.}\
  \bibnamefont {Yang}}, \bibinfo {author} {\bibfnamefont {G.~B.}\ \bibnamefont
  {Zhang}}, \ and\ \bibinfo {author} {\bibfnamefont {Y.~X.}\ \bibnamefont
  {Leng}},\ }\href {\doibase 10.1088/1361-6587/ac3c3a} {\bibfield  {journal}
  {\bibinfo  {journal} {Plasma Physics and Controlled Fusion}\ }\textbf
  {\bibinfo {volume} {64}},\ \bibinfo {pages} {025001} (\bibinfo {year}
  {2021})}\BibitemShut {NoStop}%
\bibitem [{\citenamefont {Siders}\ \emph {et~al.}(2019)\citenamefont {Siders},
  \citenamefont {Erlandson}, \citenamefont {Galvin}, \citenamefont {Frank},
  \citenamefont {Langer}, \citenamefont {Reagan}, \citenamefont {Scott},
  \citenamefont {Sistrunk},\ and\ \citenamefont {Spinka}}]{Siders2019}%
  \BibitemOpen
  \bibfield  {author} {\bibinfo {author} {\bibfnamefont {C.~W.}\ \bibnamefont
  {Siders}}, \bibinfo {author} {\bibfnamefont {A.~C.}\ \bibnamefont
  {Erlandson}}, \bibinfo {author} {\bibfnamefont {T.~C.}\ \bibnamefont
  {Galvin}}, \bibinfo {author} {\bibfnamefont {H.}~\bibnamefont {Frank}},
  \bibinfo {author} {\bibfnamefont {S.}~\bibnamefont {Langer}}, \bibinfo
  {author} {\bibfnamefont {B.~A.}\ \bibnamefont {Reagan}}, \bibinfo {author}
  {\bibfnamefont {H.}~\bibnamefont {Scott}}, \bibinfo {author} {\bibfnamefont
  {E.~F.}\ \bibnamefont {Sistrunk}}, \ and\ \bibinfo {author} {\bibfnamefont
  {T.~M.}\ \bibnamefont {Spinka}},\ }in\ \href
  {https://www.euvlitho.com/2019/S22.pdf} {\emph {\bibinfo {booktitle} {Source
  Workshop}}},\ \bibinfo {series and number} {\bibinfo {number} {S22}},\
  \bibinfo {organization} {Advanced Photon Technologies, NIF and Photon
  Science, Lawrence Livermore National Laboratory, DOE/NNSA}\ (\bibinfo
  {publisher} {EUV Litho Inc.},\ \bibinfo {year} {2019})\BibitemShut {NoStop}%
\bibitem [{\citenamefont {Fujioka}\ \emph
  {et~al.}(2005{\natexlab{b}})\citenamefont {Fujioka}, \citenamefont
  {Nishimura}, \citenamefont {Nishihara}, \citenamefont {Murakami},
  \citenamefont {Kang}, \citenamefont {Gu}, \citenamefont {Nagai},
  \citenamefont {Norimatsu}, \citenamefont {Miyanaga}, \citenamefont {Izawa},\
  and\ \citenamefont {Mima}}]{Fujioka2005}%
  \BibitemOpen
  \bibfield  {author} {\bibinfo {author} {\bibfnamefont {S.}~\bibnamefont
  {Fujioka}}, \bibinfo {author} {\bibfnamefont {H.}~\bibnamefont {Nishimura}},
  \bibinfo {author} {\bibfnamefont {K.}~\bibnamefont {Nishihara}}, \bibinfo
  {author} {\bibfnamefont {M.}~\bibnamefont {Murakami}}, \bibinfo {author}
  {\bibfnamefont {Y.-G.}\ \bibnamefont {Kang}}, \bibinfo {author}
  {\bibfnamefont {Q.}~\bibnamefont {Gu}}, \bibinfo {author} {\bibfnamefont
  {K.}~\bibnamefont {Nagai}}, \bibinfo {author} {\bibfnamefont
  {T.}~\bibnamefont {Norimatsu}}, \bibinfo {author} {\bibfnamefont
  {N.}~\bibnamefont {Miyanaga}}, \bibinfo {author} {\bibfnamefont
  {Y.}~\bibnamefont {Izawa}}, \ and\ \bibinfo {author} {\bibfnamefont
  {K.}~\bibnamefont {Mima}},\ }\href {\doibase 10.1063/1.2142102} {\bibfield
  {journal} {\bibinfo  {journal} {Appl. Phys. Lett.}\ }\textbf {\bibinfo
  {volume} {87}},\ \bibinfo {pages} {241503} (\bibinfo {year}
  {2005}{\natexlab{b}})}\BibitemShut {NoStop}%
\bibitem [{\citenamefont {Hemminga}\ \emph {et~al.}(2021)\citenamefont
  {Hemminga}, \citenamefont {Poirier}, \citenamefont {Basko}, \citenamefont
  {Hoekstra}, \citenamefont {Ubachs}, \citenamefont {Versolato},\ and\
  \citenamefont {Sheil}}]{Hemminga2021}%
  \BibitemOpen
  \bibfield  {author} {\bibinfo {author} {\bibfnamefont {D.~J.}\ \bibnamefont
  {Hemminga}}, \bibinfo {author} {\bibfnamefont {L.}~\bibnamefont {Poirier}},
  \bibinfo {author} {\bibfnamefont {M.~M.}\ \bibnamefont {Basko}}, \bibinfo
  {author} {\bibfnamefont {R.}~\bibnamefont {Hoekstra}}, \bibinfo {author}
  {\bibfnamefont {W.}~\bibnamefont {Ubachs}}, \bibinfo {author} {\bibfnamefont
  {O.~O.}\ \bibnamefont {Versolato}}, \ and\ \bibinfo {author} {\bibfnamefont
  {J.}~\bibnamefont {Sheil}},\ }\href {\doibase 10.1088/1361-6595/ac2224}
  {\bibfield  {journal} {\bibinfo  {journal} {Plasma Sources Sci. Technol.}\
  }\textbf {\bibinfo {volume} {30}},\ \bibinfo {pages} {105006} (\bibinfo
  {year} {2021})}\BibitemShut {NoStop}%
\bibitem [{\citenamefont {Schupp}\ \emph {et~al.}(2019)\citenamefont {Schupp},
  \citenamefont {Torretti}, \citenamefont {Meijer}, \citenamefont {Bayraktar},
  \citenamefont {Sheil}, \citenamefont {Scheers}, \citenamefont {Kurilovich},
  \citenamefont {Bayerle}, \citenamefont {Schafgans}, \citenamefont {Purvis},
  \citenamefont {Eikema}, \citenamefont {Witte}, \citenamefont {Ubachs},
  \citenamefont {Hoekstra},\ and\ \citenamefont {Versolato}}]{Schupp2019a}%
  \BibitemOpen
  \bibfield  {author} {\bibinfo {author} {\bibfnamefont {R.}~\bibnamefont
  {Schupp}}, \bibinfo {author} {\bibfnamefont {F.}~\bibnamefont {Torretti}},
  \bibinfo {author} {\bibfnamefont {R.~A.}\ \bibnamefont {Meijer}}, \bibinfo
  {author} {\bibfnamefont {M.}~\bibnamefont {Bayraktar}}, \bibinfo {author}
  {\bibfnamefont {J.}~\bibnamefont {Sheil}}, \bibinfo {author} {\bibfnamefont
  {J.}~\bibnamefont {Scheers}}, \bibinfo {author} {\bibfnamefont
  {D.}~\bibnamefont {Kurilovich}}, \bibinfo {author} {\bibfnamefont
  {A.}~\bibnamefont {Bayerle}}, \bibinfo {author} {\bibfnamefont {A.~A.}\
  \bibnamefont {Schafgans}}, \bibinfo {author} {\bibfnamefont {M.}~\bibnamefont
  {Purvis}}, \bibinfo {author} {\bibfnamefont {K.~S.~E.}\ \bibnamefont
  {Eikema}}, \bibinfo {author} {\bibfnamefont {S.}~\bibnamefont {Witte}},
  \bibinfo {author} {\bibfnamefont {W.}~\bibnamefont {Ubachs}}, \bibinfo
  {author} {\bibfnamefont {R.}~\bibnamefont {Hoekstra}}, \ and\ \bibinfo
  {author} {\bibfnamefont {O.~O.}\ \bibnamefont {Versolato}},\ }\href {\doibase
  10.1063/1.5117504} {\bibfield  {journal} {\bibinfo  {journal} {Appl. Phys.
  Lett.}\ }\textbf {\bibinfo {volume} {115}},\ \bibinfo {pages} {124101}
  (\bibinfo {year} {2019})}\BibitemShut {NoStop}%
\bibitem [{\citenamefont {Torretti}\ \emph {et~al.}(2019)\citenamefont
  {Torretti}, \citenamefont {Liu}, \citenamefont {Bayraktar}, \citenamefont
  {Scheers}, \citenamefont {Bouza}, \citenamefont {Ubachs}, \citenamefont
  {Hoekstra},\ and\ \citenamefont {Versolato}}]{Torretti2019}%
  \BibitemOpen
  \bibfield  {author} {\bibinfo {author} {\bibfnamefont {F.}~\bibnamefont
  {Torretti}}, \bibinfo {author} {\bibfnamefont {F.}~\bibnamefont {Liu}},
  \bibinfo {author} {\bibfnamefont {M.}~\bibnamefont {Bayraktar}}, \bibinfo
  {author} {\bibfnamefont {J.}~\bibnamefont {Scheers}}, \bibinfo {author}
  {\bibfnamefont {Z.}~\bibnamefont {Bouza}}, \bibinfo {author} {\bibfnamefont
  {W.}~\bibnamefont {Ubachs}}, \bibinfo {author} {\bibfnamefont
  {R.}~\bibnamefont {Hoekstra}}, \ and\ \bibinfo {author} {\bibfnamefont
  {O.}~\bibnamefont {Versolato}},\ }\href
  {https://doi.org/10.1088/1361-6463/ab56d4} {\bibfield  {journal} {\bibinfo
  {journal} {J. Phys. D Appl. Phys.}\ }\textbf {\bibinfo {volume} {53}},\
  \bibinfo {pages} {055204} (\bibinfo {year} {2019})}\BibitemShut {NoStop}%
\bibitem [{\citenamefont {Basko}\ \emph {et~al.}(2015)\citenamefont {Basko},
  \citenamefont {Novikov},\ and\ \citenamefont {Grushin}}]{Basko2015}%
  \BibitemOpen
  \bibfield  {author} {\bibinfo {author} {\bibfnamefont {M.~M.}\ \bibnamefont
  {Basko}}, \bibinfo {author} {\bibfnamefont {V.~G.}\ \bibnamefont {Novikov}},
  \ and\ \bibinfo {author} {\bibfnamefont {A.~S.}\ \bibnamefont {Grushin}},\
  }\href {\doibase 10.1063/1.4921334} {\bibfield  {journal} {\bibinfo
  {journal} {Phys. Plasmas}\ }\textbf {\bibinfo {volume} {22}},\ \bibinfo
  {pages} {053111} (\bibinfo {year} {2015})}\BibitemShut {NoStop}%
\bibitem [{\citenamefont {Basko}(2016)}]{Basko2016}%
  \BibitemOpen
  \bibfield  {author} {\bibinfo {author} {\bibfnamefont {M.}~\bibnamefont
  {Basko}},\ }\href {\doibase 10.1063/1.4960684} {\bibfield  {journal}
  {\bibinfo  {journal} {Phys. Plasmas}\ }\textbf {\bibinfo {volume} {23}},\
  \bibinfo {pages} {083114} (\bibinfo {year} {2016})}\BibitemShut {NoStop}%
\bibitem [{\citenamefont {Langer}\ \emph {et~al.}(2019)\citenamefont {Langer},
  \citenamefont {Siders}, \citenamefont {Galvin}, \citenamefont {Scott},\ and\
  \citenamefont {Sistrunk}}]{Langer2019}%
  \BibitemOpen
  \bibfield  {author} {\bibinfo {author} {\bibfnamefont {S.}~\bibnamefont
  {Langer}}, \bibinfo {author} {\bibfnamefont {C.}~\bibnamefont {Siders}},
  \bibinfo {author} {\bibfnamefont {T.}~\bibnamefont {Galvin}}, \bibinfo
  {author} {\bibfnamefont {H.}~\bibnamefont {Scott}}, \ and\ \bibinfo {author}
  {\bibfnamefont {E.}~\bibnamefont {Sistrunk}},\ }in\ \href
  {https://www.euvlitho.com/2019/S94.pdf} {\emph {\bibinfo {booktitle} {Source
  Workshop}}},\ \bibinfo {series and number} {\bibinfo {number} {S94}},\
  \bibinfo {organization} {Lawrence Livermore National Laboratory, DOE/NNSA}\
  (\bibinfo  {publisher} {EUV Litho Inc.},\ \bibinfo {year} {2019})\BibitemShut
  {NoStop}%
\bibitem [{\citenamefont {Fomenkov}(2019)}]{Fomenkov2019}%
  \BibitemOpen
  \bibfield  {author} {\bibinfo {author} {\bibfnamefont {I.}~\bibnamefont
  {Fomenkov}},\ }in\ \href {https://www.euvlitho.com/2019/S1.pdf} {\emph
  {\bibinfo {booktitle} {Source Workshop}}},\ \bibinfo {series and number}
  {\bibinfo {number} {S1}},\ \bibinfo {organization} {ASML}\ (\bibinfo
  {publisher} {EUV Litho Inc.},\ \bibinfo {year} {2019})\BibitemShut {NoStop}%
\bibitem [{\citenamefont {Nishihara}\ \emph {et~al.}(2006)\citenamefont
  {Nishihara}, \citenamefont {Sasaki}, \citenamefont {Sunahara},\ and\
  \citenamefont {Nishikawa}}]{Nishihara2006}%
  \BibitemOpen
  \bibfield  {author} {\bibinfo {author} {\bibfnamefont {K.}~\bibnamefont
  {Nishihara}}, \bibinfo {author} {\bibfnamefont {A.}~\bibnamefont {Sasaki}},
  \bibinfo {author} {\bibfnamefont {A.}~\bibnamefont {Sunahara}}, \ and\
  \bibinfo {author} {\bibfnamefont {T.}~\bibnamefont {Nishikawa}},\ }in\
  \href@noop {} {\emph {\bibinfo {booktitle} {EUV sources for lithography}}},\
  Vol.\ \bibinfo {volume} {149},\ \bibinfo {editor} {edited by\ \bibinfo
  {editor} {\bibfnamefont {V.}~\bibnamefont {Bakshi}}}\ (\bibinfo  {publisher}
  {SPIE Press, Bellingham, Washington},\ \bibinfo {year} {2006})\
  Chap.~\bibinfo {chapter} {11}, pp.\ \bibinfo {pages} {339 -- 370}\BibitemShut
  {NoStop}%
\bibitem [{\citenamefont {Murakami}\ \emph {et~al.}(2006)\citenamefont
  {Murakami}, \citenamefont {Fujioka}, \citenamefont {Nishimura}, \citenamefont
  {Ando}, \citenamefont {Ueda}, \citenamefont {Shimada},\ and\ \citenamefont
  {Yamaura}}]{Murakami2006}%
  \BibitemOpen
  \bibfield  {author} {\bibinfo {author} {\bibfnamefont {M.}~\bibnamefont
  {Murakami}}, \bibinfo {author} {\bibfnamefont {S.}~\bibnamefont {Fujioka}},
  \bibinfo {author} {\bibfnamefont {H.}~\bibnamefont {Nishimura}}, \bibinfo
  {author} {\bibfnamefont {T.}~\bibnamefont {Ando}}, \bibinfo {author}
  {\bibfnamefont {N.}~\bibnamefont {Ueda}}, \bibinfo {author} {\bibfnamefont
  {Y.}~\bibnamefont {Shimada}}, \ and\ \bibinfo {author} {\bibfnamefont
  {M.}~\bibnamefont {Yamaura}},\ }\href {\doibase 10.1063/1.2187445} {\bibfield
   {journal} {\bibinfo  {journal} {Phys. Plasmas}\ }\textbf {\bibinfo {volume}
  {13}},\ \bibinfo {pages} {033107} (\bibinfo {year} {2006})}\BibitemShut
  {NoStop}%
\bibitem [{\citenamefont {Basko}\ and\ \citenamefont
  {Tsygvintsev}(2017)}]{Basko2017a}%
  \BibitemOpen
  \bibfield  {author} {\bibinfo {author} {\bibfnamefont {M.~M.}\ \bibnamefont
  {Basko}}\ and\ \bibinfo {author} {\bibfnamefont {I.~P.}\ \bibnamefont
  {Tsygvintsev}},\ }\href {\doibase 10.1016/j.cpc.2017.01.010} {\bibfield
  {journal} {\bibinfo  {journal} {Comp. Phys. Comm.}\ }\textbf {\bibinfo
  {volume} {214}},\ \bibinfo {pages} {59} (\bibinfo {year} {2017})}\BibitemShut
  {NoStop}%
\bibitem [{\citenamefont {Basko}\ \emph {et~al.}(2010)\citenamefont {Basko},
  \citenamefont {Maruhn},\ and\ \citenamefont {Tauschwitz}}]{Basko2010}%
  \BibitemOpen
  \bibfield  {author} {\bibinfo {author} {\bibfnamefont {M.}~\bibnamefont
  {Basko}}, \bibinfo {author} {\bibfnamefont {J.}~\bibnamefont {Maruhn}}, \
  and\ \bibinfo {author} {\bibfnamefont {A.}~\bibnamefont {Tauschwitz}},\
  }\href@noop {} {\bibfield  {journal} {\bibinfo  {journal} {GSI report}\
  }\textbf {\bibinfo {volume} {1}},\ \bibinfo {pages} {410} (\bibinfo {year}
  {2010})}\BibitemShut {NoStop}%
\bibitem [{\citenamefont {Basko}\ \emph {et~al.}(2012)\citenamefont {Basko},
  \citenamefont {Sasorov}, \citenamefont {Murakami}, \citenamefont {Novikov},\
  and\ \citenamefont {Grushin}}]{Basko2012}%
  \BibitemOpen
  \bibfield  {author} {\bibinfo {author} {\bibfnamefont {M.~M.}\ \bibnamefont
  {Basko}}, \bibinfo {author} {\bibfnamefont {P.~V.}\ \bibnamefont {Sasorov}},
  \bibinfo {author} {\bibfnamefont {M.}~\bibnamefont {Murakami}}, \bibinfo
  {author} {\bibfnamefont {V.~G.}\ \bibnamefont {Novikov}}, \ and\ \bibinfo
  {author} {\bibfnamefont {A.~S.}\ \bibnamefont {Grushin}},\ }\href {\doibase
  10.1088/0741-3335/54/5/055003} {\bibfield  {journal} {\bibinfo  {journal}
  {Plasma Physics and Controlled Fusion}\ }\textbf {\bibinfo {volume} {54}},\
  \bibinfo {pages} {055003} (\bibinfo {year} {2012})}\BibitemShut {NoStop}%
\bibitem [{\citenamefont {Tauschwitz}\ \emph {et~al.}(2013)\citenamefont
  {Tauschwitz}, \citenamefont {Basko}, \citenamefont {Frank}, \citenamefont
  {Novikov}, \citenamefont {Grushin}, \citenamefont {Blazevic}, \citenamefont
  {Roth},\ and\ \citenamefont {Maruhn}}]{Tauschwitz2013}%
  \BibitemOpen
  \bibfield  {author} {\bibinfo {author} {\bibfnamefont {A.}~\bibnamefont
  {Tauschwitz}}, \bibinfo {author} {\bibfnamefont {M.}~\bibnamefont {Basko}},
  \bibinfo {author} {\bibfnamefont {A.}~\bibnamefont {Frank}}, \bibinfo
  {author} {\bibfnamefont {V.}~\bibnamefont {Novikov}}, \bibinfo {author}
  {\bibfnamefont {A.}~\bibnamefont {Grushin}}, \bibinfo {author} {\bibfnamefont
  {A.}~\bibnamefont {Blazevic}}, \bibinfo {author} {\bibfnamefont
  {M.}~\bibnamefont {Roth}}, \ and\ \bibinfo {author} {\bibfnamefont
  {J.}~\bibnamefont {Maruhn}},\ }\href {\doibase
  https://doi.org/10.1016/j.hedp.2012.12.004} {\bibfield  {journal} {\bibinfo
  {journal} {High Energy Density Phys.}\ }\textbf {\bibinfo {volume} {9}},\
  \bibinfo {pages} {158} (\bibinfo {year} {2013})}\BibitemShut {NoStop}%
\bibitem [{\citenamefont {Basko}\ \emph {et~al.}(2017)\citenamefont {Basko},
  \citenamefont {Krivokorytov}, \citenamefont {Vinokhodov}, \citenamefont
  {Sidelnikov}, \citenamefont {Krivtsun}, \citenamefont {Medvedev},
  \citenamefont {Kim}, \citenamefont {Kompanets}, \citenamefont {Lash},\ and\
  \citenamefont {Koshelev}}]{Basko2017b}%
  \BibitemOpen
  \bibfield  {author} {\bibinfo {author} {\bibfnamefont {M.~M.}\ \bibnamefont
  {Basko}}, \bibinfo {author} {\bibfnamefont {M.~S.}\ \bibnamefont
  {Krivokorytov}}, \bibinfo {author} {\bibfnamefont {A.~Y.}\ \bibnamefont
  {Vinokhodov}}, \bibinfo {author} {\bibfnamefont {Y.~V.}\ \bibnamefont
  {Sidelnikov}}, \bibinfo {author} {\bibfnamefont {V.~M.}\ \bibnamefont
  {Krivtsun}}, \bibinfo {author} {\bibfnamefont {V.~V.}\ \bibnamefont
  {Medvedev}}, \bibinfo {author} {\bibfnamefont {D.~A.}\ \bibnamefont {Kim}},
  \bibinfo {author} {\bibfnamefont {V.~O.}\ \bibnamefont {Kompanets}}, \bibinfo
  {author} {\bibfnamefont {A.~A.}\ \bibnamefont {Lash}}, \ and\ \bibinfo
  {author} {\bibfnamefont {K.~N.}\ \bibnamefont {Koshelev}},\ }\href {\doibase
  10.1088/1612-202x/aa539b} {\bibfield  {journal} {\bibinfo  {journal} {Laser.
  Phys. Lett.}\ }\textbf {\bibinfo {volume} {14}},\ \bibinfo {pages} {036001}
  (\bibinfo {year} {2017})}\BibitemShut {NoStop}%
\bibitem [{\citenamefont {Kurilovich}\ \emph {et~al.}(2018)\citenamefont
  {Kurilovich}, \citenamefont {Basko}, \citenamefont {Kim}, \citenamefont
  {Torretti}, \citenamefont {Schupp}, \citenamefont {Visschers}, \citenamefont
  {Scheers}, \citenamefont {Hoekstra}, \citenamefont {Ubachs},\ and\
  \citenamefont {Versolato}}]{Kurilovich2018}%
  \BibitemOpen
  \bibfield  {author} {\bibinfo {author} {\bibfnamefont {D.}~\bibnamefont
  {Kurilovich}}, \bibinfo {author} {\bibfnamefont {M.~M.}\ \bibnamefont
  {Basko}}, \bibinfo {author} {\bibfnamefont {D.~A.}\ \bibnamefont {Kim}},
  \bibinfo {author} {\bibfnamefont {F.}~\bibnamefont {Torretti}}, \bibinfo
  {author} {\bibfnamefont {R.}~\bibnamefont {Schupp}}, \bibinfo {author}
  {\bibfnamefont {J.~C.}\ \bibnamefont {Visschers}}, \bibinfo {author}
  {\bibfnamefont {J.}~\bibnamefont {Scheers}}, \bibinfo {author} {\bibfnamefont
  {R.}~\bibnamefont {Hoekstra}}, \bibinfo {author} {\bibfnamefont
  {W.}~\bibnamefont {Ubachs}}, \ and\ \bibinfo {author} {\bibfnamefont {O.~O.}\
  \bibnamefont {Versolato}},\ }\href {\doibase 10.1063/1.5010899} {\bibfield
  {journal} {\bibinfo  {journal} {Phys. Plasmas}\ }\textbf {\bibinfo {volume}
  {25}},\ \bibinfo {pages} {012709} (\bibinfo {year} {2018})}\BibitemShut
  {NoStop}%
\bibitem [{\citenamefont {Basko}\ \emph {et~al.}(2009)\citenamefont {Basko},
  \citenamefont {Maruhn},\ and\ \citenamefont {Tauschwitz}}]{Basko2009}%
  \BibitemOpen
  \bibfield  {author} {\bibinfo {author} {\bibfnamefont {M.~M.}\ \bibnamefont
  {Basko}}, \bibinfo {author} {\bibfnamefont {J.}~\bibnamefont {Maruhn}}, \
  and\ \bibinfo {author} {\bibfnamefont {A.}~\bibnamefont {Tauschwitz}},\
  }\href {https://doi.org/10.1016/j.jcp.2008.11.031} {\bibfield  {journal}
  {\bibinfo  {journal} {Journal of Computational Physics}\ }\textbf {\bibinfo
  {volume} {228}},\ \bibinfo {pages} {2175} (\bibinfo {year}
  {2009})}\BibitemShut {NoStop}%
\bibitem [{\citenamefont {Nikiforov}\ \emph {et~al.}(2005)\citenamefont
  {Nikiforov}, \citenamefont {Novikov},\ and\ \citenamefont
  {Uvarov}}]{Nikiforov2005}%
  \BibitemOpen
  \bibfield  {author} {\bibinfo {author} {\bibfnamefont {A.~F.}\ \bibnamefont
  {Nikiforov}}, \bibinfo {author} {\bibfnamefont {V.~G.}\ \bibnamefont
  {Novikov}}, \ and\ \bibinfo {author} {\bibfnamefont {V.~B.}\ \bibnamefont
  {Uvarov}},\ }\href@noop {} {\emph {\bibinfo {title} {{Quantum-Statistical
  Models of Hot Dense Matter: Methods for Computation Opacity and Equation of
  State (Progress in Mathematical Physics)}}}}\ (\bibinfo  {publisher}
  {Birkhauser},\ \bibinfo {year} {2005})\BibitemShut {NoStop}%
\bibitem [{\citenamefont {Vichev}\ \emph {et~al.}(2019)\citenamefont {Vichev},
  \citenamefont {Solomyannaya}, \citenamefont {Grushin},\ and\ \citenamefont
  {Kim}}]{Vichev2019}%
  \BibitemOpen
  \bibfield  {author} {\bibinfo {author} {\bibfnamefont {I.}~\bibnamefont
  {Vichev}}, \bibinfo {author} {\bibfnamefont {A.}~\bibnamefont
  {Solomyannaya}}, \bibinfo {author} {\bibfnamefont {A.}~\bibnamefont
  {Grushin}}, \ and\ \bibinfo {author} {\bibfnamefont {D.}~\bibnamefont
  {Kim}},\ }\href {\doibase https://doi.org/10.1016/j.hedp.2019.100713}
  {\bibfield  {journal} {\bibinfo  {journal} {High Energy Density Physics}\
  }\textbf {\bibinfo {volume} {33}},\ \bibinfo {pages} {100713} (\bibinfo
  {year} {2019})}\BibitemShut {NoStop}%
\bibitem [{\citenamefont {Faik}\ \emph {et~al.}(2018)\citenamefont {Faik},
  \citenamefont {Tauschwitz},\ and\ \citenamefont {Iosilevskiy}}]{Faik2018}%
  \BibitemOpen
  \bibfield  {author} {\bibinfo {author} {\bibfnamefont {S.}~\bibnamefont
  {Faik}}, \bibinfo {author} {\bibfnamefont {A.}~\bibnamefont {Tauschwitz}}, \
  and\ \bibinfo {author} {\bibfnamefont {I.}~\bibnamefont {Iosilevskiy}},\
  }\href {\doibase https://doi.org/10.1016/j.cpc.2018.01.008} {\bibfield
  {journal} {\bibinfo  {journal} {Comp. Phys. Comm.}\ }\textbf {\bibinfo
  {volume} {227}},\ \bibinfo {pages} {117} (\bibinfo {year}
  {2018})}\BibitemShut {NoStop}%
\bibitem [{\citenamefont {Elezier}(2002)}]{Elezier2002}%
  \BibitemOpen
  \bibfield  {author} {\bibinfo {author} {\bibfnamefont {S.}~\bibnamefont
  {Elezier}},\ }\href@noop {} {\emph {\bibinfo {title} {The Interaction of
  High-Power Lasers with Plasmas}}}\ (\bibinfo  {publisher} {CRC Press},\
  \bibinfo {year} {2002})\BibitemShut {NoStop}%
\bibitem [{\citenamefont {{R. Schupp, F. Torretti, R. A. Meijer, M. Bayraktar,
  J. Scheers, D. Kurilovich, A. Bayerle, K. S. E. Eikema, S. Witte, W. Ubachs,
  R. Hoekstra and O. O. Versolato}}(2019)}]{Schupp2019}%
  \BibitemOpen
  \bibfield  {author} {\bibinfo {author} {\bibnamefont {{R. Schupp, F.
  Torretti, R. A. Meijer, M. Bayraktar, J. Scheers, D. Kurilovich, A. Bayerle,
  K. S. E. Eikema, S. Witte, W. Ubachs, R. Hoekstra and O. O. Versolato}}},\
  }\href {\doibase https://doi.org/10.1103/PhysRevApplied.12.014010} {\bibfield
   {journal} {\bibinfo  {journal} {Phys. Rev. Appl.}\ }\textbf {\bibinfo
  {volume} {12}},\ \bibinfo {pages} {014010} (\bibinfo {year}
  {2019})}\BibitemShut {NoStop}%
\bibitem [{\citenamefont {Schupp}\ \emph
  {et~al.}(2021{\natexlab{b}})\citenamefont {Schupp}, \citenamefont {Behnke},
  \citenamefont {Bouza}, \citenamefont {Mazzotta}, \citenamefont {Mostafa},
  \citenamefont {Lassise}, \citenamefont {Poirier}, \citenamefont {Sheil},
  \citenamefont {Bayraktar}, \citenamefont {Ubachs}, \citenamefont {Hoekstra},\
  and\ \citenamefont {Versolato}}]{Schupp2021a}%
  \BibitemOpen
  \bibfield  {author} {\bibinfo {author} {\bibfnamefont {R.}~\bibnamefont
  {Schupp}}, \bibinfo {author} {\bibfnamefont {L.}~\bibnamefont {Behnke}},
  \bibinfo {author} {\bibfnamefont {Z.}~\bibnamefont {Bouza}}, \bibinfo
  {author} {\bibfnamefont {Z.}~\bibnamefont {Mazzotta}}, \bibinfo {author}
  {\bibfnamefont {Y.}~\bibnamefont {Mostafa}}, \bibinfo {author} {\bibfnamefont
  {A.}~\bibnamefont {Lassise}}, \bibinfo {author} {\bibfnamefont
  {L.}~\bibnamefont {Poirier}}, \bibinfo {author} {\bibfnamefont
  {J.}~\bibnamefont {Sheil}}, \bibinfo {author} {\bibfnamefont
  {M.}~\bibnamefont {Bayraktar}}, \bibinfo {author} {\bibfnamefont
  {W.}~\bibnamefont {Ubachs}}, \bibinfo {author} {\bibfnamefont
  {R.}~\bibnamefont {Hoekstra}}, \ and\ \bibinfo {author} {\bibfnamefont
  {O.~O.}\ \bibnamefont {Versolato}},\ }\href {\doibase
  10.1088/1361-6463/ac0b70} {\bibfield  {journal} {\bibinfo  {journal} {Journal
  of Physics D: Applied Physics}\ }\textbf {\bibinfo {volume} {54}},\ \bibinfo
  {pages} {365103} (\bibinfo {year} {2021}{\natexlab{b}})}\BibitemShut
  {NoStop}%
\bibitem [{\citenamefont {Shimada}\ \emph {et~al.}(2005)\citenamefont
  {Shimada}, \citenamefont {Nishimura}, \citenamefont {Nakai}, \citenamefont
  {Hashimoto}, \citenamefont {Yamaura}, \citenamefont {Tao}, \citenamefont
  {Shigemori}, \citenamefont {Okuno}, \citenamefont {Nishihara}, \citenamefont
  {Kawamura} \emph {et~al.}}]{Shimada2005}%
  \BibitemOpen
  \bibfield  {author} {\bibinfo {author} {\bibfnamefont {Y.}~\bibnamefont
  {Shimada}}, \bibinfo {author} {\bibfnamefont {H.}~\bibnamefont {Nishimura}},
  \bibinfo {author} {\bibfnamefont {M.}~\bibnamefont {Nakai}}, \bibinfo
  {author} {\bibfnamefont {K.}~\bibnamefont {Hashimoto}}, \bibinfo {author}
  {\bibfnamefont {M.}~\bibnamefont {Yamaura}}, \bibinfo {author} {\bibfnamefont
  {Y.}~\bibnamefont {Tao}}, \bibinfo {author} {\bibfnamefont {K.}~\bibnamefont
  {Shigemori}}, \bibinfo {author} {\bibfnamefont {T.}~\bibnamefont {Okuno}},
  \bibinfo {author} {\bibfnamefont {K.}~\bibnamefont {Nishihara}}, \bibinfo
  {author} {\bibfnamefont {T.}~\bibnamefont {Kawamura}},  \emph {et~al.},\
  }\href {\doibase 10.1063/1.1856697} {\bibfield  {journal} {\bibinfo
  {journal} {Appl. Phys. Lett.}\ }\textbf {\bibinfo {volume} {86}},\ \bibinfo
  {pages} {051501} (\bibinfo {year} {2005})}\BibitemShut {NoStop}%
\bibitem [{\citenamefont {Giovannini}\ and\ \citenamefont
  {Abhari}(2013)}]{Giovannini2013}%
  \BibitemOpen
  \bibfield  {author} {\bibinfo {author} {\bibfnamefont {A.~Z.}\ \bibnamefont
  {Giovannini}}\ and\ \bibinfo {author} {\bibfnamefont {R.~S.}\ \bibnamefont
  {Abhari}},\ }\href {\doibase 10.1063/1.4815955} {\bibfield  {journal}
  {\bibinfo  {journal} {J. Appl. Phys.}\ }\textbf {\bibinfo {volume} {114}},\
  \bibinfo {pages} {033303} (\bibinfo {year} {2013})}\BibitemShut {NoStop}%
\bibitem [{\citenamefont {Giovannini}\ and\ \citenamefont
  {Abhari}(2014)}]{Giovannini2014}%
  \BibitemOpen
  \bibfield  {author} {\bibinfo {author} {\bibfnamefont {A.~Z.}\ \bibnamefont
  {Giovannini}}\ and\ \bibinfo {author} {\bibfnamefont {R.~S.}\ \bibnamefont
  {Abhari}},\ }\href {\doibase 10.1063/1.4878506} {\bibfield  {journal}
  {\bibinfo  {journal} {Appl. Phys. Lett.}\ }\textbf {\bibinfo {volume}
  {104}},\ \bibinfo {pages} {194104} (\bibinfo {year} {2014})}\BibitemShut
  {NoStop}%
\bibitem [{\citenamefont {Giovannini}\ \emph {et~al.}(2015)\citenamefont
  {Giovannini}, \citenamefont {Gambino}, \citenamefont {Rollinger},\ and\
  \citenamefont {Abhari}}]{Giovannini2015}%
  \BibitemOpen
  \bibfield  {author} {\bibinfo {author} {\bibfnamefont {A.~Z.}\ \bibnamefont
  {Giovannini}}, \bibinfo {author} {\bibfnamefont {N.}~\bibnamefont {Gambino}},
  \bibinfo {author} {\bibfnamefont {B.}~\bibnamefont {Rollinger}}, \ and\
  \bibinfo {author} {\bibfnamefont {R.~S.}\ \bibnamefont {Abhari}},\
  }\href@noop {} {\bibfield  {journal} {\bibinfo  {journal} {J. Appl. Phys.}\
  }\textbf {\bibinfo {volume} {117}},\ \bibinfo {pages} {033302} (\bibinfo
  {year} {2015})}\BibitemShut {NoStop}%
\bibitem [{\citenamefont {Sizyuk}\ \emph {et~al.}(2006)\citenamefont {Sizyuk},
  \citenamefont {Hassanein},\ and\ \citenamefont {Sizyuk}}]{Sizyuk2006}%
  \BibitemOpen
  \bibfield  {author} {\bibinfo {author} {\bibfnamefont {V.}~\bibnamefont
  {Sizyuk}}, \bibinfo {author} {\bibfnamefont {A.}~\bibnamefont {Hassanein}}, \
  and\ \bibinfo {author} {\bibfnamefont {T.}~\bibnamefont {Sizyuk}},\
  }\href@noop {} {\bibfield  {journal} {\bibinfo  {journal} {Journal of Applied
  Physics}\ }\textbf {\bibinfo {volume} {100}},\ \bibinfo {pages} {103106}
  (\bibinfo {year} {2006})}\BibitemShut {NoStop}%
\bibitem [{\citenamefont {Behnke}\ \emph {et~al.}(2021)\citenamefont {Behnke},
  \citenamefont {Schupp}, \citenamefont {Bouza}, \citenamefont {Bayraktar},
  \citenamefont {Mazzotta}, \citenamefont {Meijer}, \citenamefont {Sheil},
  \citenamefont {Witte}, \citenamefont {Ubachs}, \citenamefont {Hoekstra},\
  and\ \citenamefont {Versolato}}]{Behnke2021}%
  \BibitemOpen
  \bibfield  {author} {\bibinfo {author} {\bibfnamefont {L.}~\bibnamefont
  {Behnke}}, \bibinfo {author} {\bibfnamefont {R.}~\bibnamefont {Schupp}},
  \bibinfo {author} {\bibfnamefont {Z.}~\bibnamefont {Bouza}}, \bibinfo
  {author} {\bibfnamefont {M.}~\bibnamefont {Bayraktar}}, \bibinfo {author}
  {\bibfnamefont {Z.}~\bibnamefont {Mazzotta}}, \bibinfo {author}
  {\bibfnamefont {R.}~\bibnamefont {Meijer}}, \bibinfo {author} {\bibfnamefont
  {J.}~\bibnamefont {Sheil}}, \bibinfo {author} {\bibfnamefont
  {S.}~\bibnamefont {Witte}}, \bibinfo {author} {\bibfnamefont
  {W.}~\bibnamefont {Ubachs}}, \bibinfo {author} {\bibfnamefont
  {R.}~\bibnamefont {Hoekstra}}, \ and\ \bibinfo {author} {\bibfnamefont
  {O.~O.}\ \bibnamefont {Versolato}},\ }\href {\doibase 10.1364/OE.411539}
  {\bibfield  {journal} {\bibinfo  {journal} {Opt. Express}\ }\textbf {\bibinfo
  {volume} {29}},\ \bibinfo {pages} {4475} (\bibinfo {year}
  {2021})}\BibitemShut {NoStop}%
\bibitem [{\citenamefont {Nishihara}(1982)}]{Nishihara1982}%
  \BibitemOpen
  \bibfield  {author} {\bibinfo {author} {\bibfnamefont {K.}~\bibnamefont
  {Nishihara}},\ }\href {\doibase 10.1143/jjap.21.l571} {\bibfield  {journal}
  {\bibinfo  {journal} {Japanese Journal of Applied Physics}\ }\textbf
  {\bibinfo {volume} {21}},\ \bibinfo {pages} {L571} (\bibinfo {year}
  {1982})}\BibitemShut {NoStop}%
\bibitem [{\citenamefont {Tanaka}\ \emph {et~al.}(2015)\citenamefont {Tanaka},
  \citenamefont {Masuda}, \citenamefont {Deguchi}, \citenamefont {Murakami},
  \citenamefont {Sunahara}, \citenamefont {Fujioka}, \citenamefont {Yogo},\
  and\ \citenamefont {Nishimura}}]{Nozomi2015}%
  \BibitemOpen
  \bibfield  {author} {\bibinfo {author} {\bibfnamefont {N.}~\bibnamefont
  {Tanaka}}, \bibinfo {author} {\bibfnamefont {M.}~\bibnamefont {Masuda}},
  \bibinfo {author} {\bibfnamefont {R.}~\bibnamefont {Deguchi}}, \bibinfo
  {author} {\bibfnamefont {M.}~\bibnamefont {Murakami}}, \bibinfo {author}
  {\bibfnamefont {A.}~\bibnamefont {Sunahara}}, \bibinfo {author}
  {\bibfnamefont {S.}~\bibnamefont {Fujioka}}, \bibinfo {author} {\bibfnamefont
  {A.}~\bibnamefont {Yogo}}, \ and\ \bibinfo {author} {\bibfnamefont
  {H.}~\bibnamefont {Nishimura}},\ }\href {\doibase 10.1063/1.4930958}
  {\bibfield  {journal} {\bibinfo  {journal} {Applied Physics Letters}\
  }\textbf {\bibinfo {volume} {107}},\ \bibinfo {pages} {114101} (\bibinfo
  {year} {2015})},\ \Eprint
  {http://arxiv.org/abs/https://doi.org/10.1063/1.4930958}
  {https://doi.org/10.1063/1.4930958} \BibitemShut {NoStop}%
\bibitem [{\citenamefont {Faik}\ \emph {et~al.}(2014)\citenamefont {Faik},
  \citenamefont {Tauschwitz}, \citenamefont {Basko}, \citenamefont {Maruhn},
  \citenamefont {Rosmej}, \citenamefont {Rienecker}, \citenamefont {Novikov},\
  and\ \citenamefont {Grushin}}]{Faik2014}%
  \BibitemOpen
  \bibfield  {author} {\bibinfo {author} {\bibfnamefont {S.}~\bibnamefont
  {Faik}}, \bibinfo {author} {\bibfnamefont {A.}~\bibnamefont {Tauschwitz}},
  \bibinfo {author} {\bibfnamefont {M.~M.}\ \bibnamefont {Basko}}, \bibinfo
  {author} {\bibfnamefont {J.~A.}\ \bibnamefont {Maruhn}}, \bibinfo {author}
  {\bibfnamefont {O.}~\bibnamefont {Rosmej}}, \bibinfo {author} {\bibfnamefont
  {T.}~\bibnamefont {Rienecker}}, \bibinfo {author} {\bibfnamefont {V.~G.}\
  \bibnamefont {Novikov}}, \ and\ \bibinfo {author} {\bibfnamefont {A.~S.}\
  \bibnamefont {Grushin}},\ }\href {\doibase
  https://doi.org/10.1016/j.hedp.2013.10.002} {\bibfield  {journal} {\bibinfo
  {journal} {High Energy Density Phys.}\ }\textbf {\bibinfo {volume} {10}},\
  \bibinfo {pages} {47 } (\bibinfo {year} {2014})}\BibitemShut {NoStop}%
\bibitem [{\citenamefont {Pütterich}\ \emph {et~al.}(2008)\citenamefont
  {Pütterich}, \citenamefont {Neu}, \citenamefont {Dux}, \citenamefont
  {Whiteford},\ and\ \citenamefont {and}}]{Putterich2008}%
  \BibitemOpen
  \bibfield  {author} {\bibinfo {author} {\bibfnamefont {T.}~\bibnamefont
  {Pütterich}}, \bibinfo {author} {\bibfnamefont {R.}~\bibnamefont {Neu}},
  \bibinfo {author} {\bibfnamefont {R.}~\bibnamefont {Dux}}, \bibinfo {author}
  {\bibfnamefont {A.~D.}\ \bibnamefont {Whiteford}}, \ and\ \bibinfo {author}
  {\bibfnamefont {M.~G.~O.}\ \bibnamefont {and}},\ }\href {\doibase
  10.1088/0741-3335/50/8/085016} {\bibfield  {journal} {\bibinfo  {journal}
  {Plasma Physics and Controlled Fusion}\ }\textbf {\bibinfo {volume} {50}},\
  \bibinfo {pages} {085016} (\bibinfo {year} {2008})}\BibitemShut {NoStop}%
\end{thebibliography}%

\end{document}